\documentclass[fleqn,usenatbib]{mnras}

\usepackage{newtxtext,newtxmath}

\usepackage[T1]{fontenc}
\usepackage{soul}

\usepackage[dvipsnames]{xcolor}
\usepackage{graphicx}	
\usepackage{amsmath}	
\usepackage{tikz}
\usetikzlibrary{shapes.geometric}
\usepackage{bm}
\usepackage[normalem]{ulem}

\usepackage{xcolor}
\DeclareUnicodeCharacter{0002}{\textcolor{red}{BAD!!}}

\newcommand{\proba}{P}
\definecolor{darkgreen}{rgb}{0.0,0.5,0.0}

\newcommand{\Mpch}{\ensuremath{h^{-1}\;\text{Mpc}}}
\newcommand{\hMpc}{\ensuremath{h\;\text{Mpc}^{-1}}}
\newcommand{\kms}{\ensuremath{\mathrm{km}\;\mathrm{s}^{-1}}}

\usepackage{hyperref}

\title[Physical Inference From Peculiar Velocity Tracers]{Field-Based Physical Inference From Peculiar Velocity Tracers}

\author[J. Prideaux-Ghee et al.]{
James Prideaux-Ghee,$^{1}$\thanks{E-mail: j.prideaux-ghee19@imperial.ac.uk (JPG)}
Florent Leclercq,$^{2,1}$ 
Guilhem Lavaux,$^{2}$ 
Alan Heavens,$^{1}$ 
and Jens Jasche$^{3}$\\
$^{1}$Imperial Centre for Inference and Cosmology (ICIC) \& Astrophysics group, Imperial College, Blackett Laboratory,
Prince Consort Road, London SW7 2AZ, UK\\
$^{2}$CNRS \& Sorbonne Université, UMR 7095, Institut d’Astrophysique de Paris, 98 bis boulevard Arago, F-75014 Paris, France\\
$^{3}$The Oskar Klein Centre, Department of Physics, Stockholm University, Albanova University Center, SE-106 91 Stockholm, Sweden\\
}

\date{Accepted XXX. Received YYY; in original form ZZZ}

\pubyear{2021}

\begin{document}
\label{firstpage}
\pagerange{\pageref{firstpage}--\pageref{lastpage}}
\maketitle

\begin{abstract}
We present \textcolor{black}{a proof-of-concept} Bayesian hierarchical modelling approach to reconstruct the initial cosmic matter density field constrained by peculiar velocity observations. Using a model for the gravitational evolution of dark matter to connect the initial conditions to late-time observations, it reconstructs the late-time density and velocity fields as natural byproducts.
We implement this field-based physical inference approach by adapting the Bayesian Origin Reconstruction from Galaxies ($\textsc{borg}$) algorithm, which explores the high-dimensional posterior through the use of Hamiltonian Monte Carlo sampling. We test the self-consistency of the method using random sets of tracers, and assess its accuracy in a more complex scenario where peculiar velocity tracers are mock haloes drawn from $\textsc{gadget2}$ N-body simulations. 
We find that our framework self-consistently infers the initial conditions, density and velocity fields, and shows some robustness to model mis-specification. Compared with the 
approach of constrained Gaussian random fields/Wiener filtering, the hierarchical model produces more accurate final density and velocity field reconstructions. It also allows us to constrain the initial conditions by peculiar velocity observations, complementing in this aspect \textcolor{black}{other field-based approaches based on alternative cosmological observables such as galaxy clustering or weak lensing}.

\end{abstract}




\section{Introduction}

The gravitational interaction of galaxies with the underlying cosmic matter density field introduces velocities in addition to the universal expansion. These peculiar velocities are assumed to trace the matter peculiar velocity field, which is dynamically linked to the underlying matter density field. Radial peculiar velocities can be measured for nearby galaxies with independent distance modulus estimates. Galaxies with distance indicators are therefore a tool for describing both the large-scale matter distribution and its dynamics in the local Universe. Since the late-time matter distribution is dependent on primordial matter perturbations and on cosmological parameters, it can be used to study the perturbations and constrain these parameters \citep{1997ApJ...486...21Z,2017MNRAS.471.3135H,2018MNRAS.480.5332W,2020MNRAS.498.2703B}.

\textcolor{black}{Furthermore, galaxies with distance indicators are rare probes of the nearby gravitational potential, since there are few nearby galaxy clusters with enough background sources to provide good weak lensing observations \citep{2015MNRAS.450.3665S}. Current mass estimates for these clusters are found to differ from estimates found via the Sunyaev-Zel'dovich effect \citep[see e.g.][]{2021MNRAS.507.5425S}. Differences in the estimated masses between various methods lead to different inferred values of the cosmological parameter $\sigma_{8}$. This has resulted in the so-called `$\sigma_{8}$ tension' -- a few percent-level tension in the value of $\sigma_{8}$ between the Planck value and values from other sources, such as the DES or KiDS surveys \citep[see][and references therein]{2020A&A...641A...6P}. Local galaxies with distance indicators could therefore be used to improve the estimates of the dynamical masses of local galaxy clusters. This requires the assumption of a $\Lambda$CDM cosmology (which we do throughout this work), such that the gravitational slip is zero \citep{2011RSPTA.369.4947B} and so galaxy static and dynamic masses are equivalent \citep{2019MNRAS.486..596P}.
}

Catalogues containing thousands of peculiar velocity tracers are now available, for example the Cosmicflows-4 catalogue \citep{2020ApJ...902..145K,2022arXiv220911238T}, and the 6dF galaxy survey \citep{2014MNRAS.443.1231C}. Distance moduli estimates are not available for every nearby galaxy with a measured redshift, so peculiar velocity data sets are usually sparser than galaxy redshift surveys. Statistical methods allow us to use such limited data sets as the basis for inferring the local peculiar velocity field, which we can then use as an additional probe of cosmology. \textcolor{black}{A significant challenge is that the statistical properties of the late-time peculiar velocity field are not not known in a closed-form algebraic description.}

\textcolor{black}{Several methods have been implemented to perform reconstructions of the velocity field. Early methods include $\textsc{potent}$  \citep{1989ApJ...336L...5B,1990ApJ...364..349D,1999ApJ...522....1D} and $\textsc{velmod}$ \citep{1997ApJ...486..629W, 1998ApJ...507...64W}. Under the assumptions of the standard cosmological framework, the continuity equation for dark matter and, crucially, linear dynamics,  the matter density and velocity field are both well-described by a Gaussian Random Field (GRF), motivating Wiener filtering reconstructions  \citep[e.g.][]{1999ApJ...520..413Z, 2012ApJ...744...43C,2014Natur.513...71T, 2015MNRAS.449.4494H} using Cosmicflows catalogues \citep{2008ApJ...676..184T,2013AJ....146...86T}. Assuming linear dynamics, constrained Gaussian realisations of the initial conditions can also be obtained \citep[e.g.][]{Hoffman1991,1996MNRAS.281...84V,10.1111/j.1365-2966.2008.13341.x,YEPES20141,Sorce2016,2018NatAs...2..680H,10.1093/mnras/staa2541}. More recently, deep learning methods have been employed, for example \citet{2021ApJ...913...76H} learned the velocity field from the Cosmicflows-3 catalogue \citep{2016AJ....152...50T}.} 

\textcolor{black}{$\textsc{virbius}$ \citep{2016MNRAS.457..172L} went beyond the Wiener filtering approach by performing Bayesian inference of the velocity field using Gaussian likelihoods for the field, the measured distance moduli, and redshifts of tracers.} $\textsc{virbius}$ accounts for the uncertainty in the measured distance moduli and redshifts of tracers, in order to account for 
\textcolor{black}{biases inherent to peculiar velocity data.}
\textsc{virbius} combined reconstructions with cosmological parameter estimation and distance sampling, producing reconstructions of the velocity field as well as sampling cosmological parameters such as the Hubble constant and power spectrum amplitude. The algorithm was tested using GRFs and halo-based mock catalogues.

\textcolor{black}{The state-of-the-art forward modelling approaches of \citet{2019MNRAS.488.5438G} and \citet{2022MNRAS.513.5148V} improve upon $\textsc{virbius}$ and reconstruct the late-time velocity and density contrast fields from late-time observations of galaxies with distance indicators. Both assume a Gaussian prior on the late-time velocity field, which is the maximum entropy prior when only the mean and covariance are known \citep[e.g.][]{jaynes03}. Both of these schemes account for any non-linearity present in the data by broadening their likelihood with an additional variance term denoted $\sigma_{\mathrm{NL}}$, as in $\textsc{virbius}$. The presence of $\sigma_{\mathrm{NL}}$ broadens the posterior distributions for quantities such as the matter power spectrum, allowing these methods to produce unbiased inferences. Both also model the distances to galaxy tracers in order to account for homogeneous Malmquist bias (see section \ref{PVTracers} for further discussion). \citet{2019MNRAS.488.5438G} applied their method to the Cosmicflows-3 catalogue.}

\textcolor{black}{Other methods to minimise the biases present in peculiar velocity catalogues have been employed. For example, \citet{Sorce2015} aimed to minimise the Malmquist bias and lognormal error in the Cosmicflows-2 catalogue by applying corrections to the observed tracer peculiar velocities, in order to make their distribution Gaussian. \citet{Hoffman2021} derived Gaussian-distributed estimators for galaxy distances and peculiar velocities in order to remove lognormal bias from data sets. Both methods were tested using the Wiener filter/constrained realisations approach.}

\textcolor{black}{Reconstruction schemes exist to attempt to reconstruct the cosmic dark matter initial conditions, from which constrained realisations of the late-time fields can be constructed. The known Gaussian statistics of the initial conditions \citep{2020A&A...641A...9P} are made use of in many such schemes. These schemes can generally be split into backwards and forwards modelling approaches, depending on whether the method draws samples of the late-time fields and rewinds to the initial conditions, or samples the initial conditions and evolves them to the late-time fields. Examples of backwards modelling approaches include the approach of \citet{1996ApJ...458..419K}, and the Reverse Zel'dovich Approximation \citep{2013MNRAS.430..888D,2013MNRAS.430..902D,2013MNRAS.430..912D}, which has been applied to the Cosmicflows-1 and Cosmicflows-2 catalogues \citep{2014MNRAS.437.3586S,10.1093/mnras/sty505,2020MNRAS.496.5139S}. Examples of forwards modelling approaches include Augmented Lagrangian Perturbation Theory applied to the 2MASS catalogue \citep{2013MNRAS.435.2065H} or the \textsc{cosmic birth} algorithm \citep{2021MNRAS.502.3456K,2021MNRAS.500.3194A}, and the \textsc{elucid} algorithm \citep{2014ApJ...794...94W,2016ApJ...831..164W,2017ApJ...841...55T}.
}

\textcolor{black}{The $\textsc{borg}$ algorithm \citep{2013MNRAS.432..894J,2015JCAP...01..036J,2019A&A...625A..64J} produces Bayesian posterior distributions for the discretised initial conditions using a forward modelling approach, originally conditioned on galaxy clustering data \citep{2015JCAP...01..036J,2016MNRAS.455.3169L,2019arXiv190906396L}. Part of the power of $\textsc{borg}$ is its modularity: the algorithm has recently been extended to cosmic shear data \citep{2021MNRAS.502.3035P,2022MNRAS.509.3194P}. The initial conditions samples $\textsc{borg}$ produces can also be used to construct the velocity field \citep{2015JCAP...06..015L,2017JCAP...06..049L}. Constrained realisations of the late-time density contrast produced from $\textsc{borg}$ initial conditions have been applied to a number of situations  \citep[e.g.][]{PhysRevD.98.083010,PhysRevD.103.023523,2022JCAP...08..003T}. The initial density and late-time observations are linked through the use of a structure formation model, for example, Lagrangian Perturbation Theory (LPT), which accurately describes linear and mildly non-linear regimes \citep{2002PhR...367....1B}. Such a data model allows for a more physical representation of the data than if we assume that the dynamics is described by first-order Eulerian perturbation theory.}

\textcolor{black}{In this work, we build upon the existing $\textsc{borg}$ machinery and infer the initial density field conditioned on peculiar velocity tracer data, using LPT to link the two, and without having to deal with the complexities of galaxy bias. As part of the inference, we obtain as natural byproducts samples of the peculiar velocity and density contrast at redshift zero (and can generate these at any redshift if desired).}

The paper is organised as follows. In section \ref{sec::method} we present the method. In particular, we discuss our modelling assumptions, the data model used in the reconstruction method, and then the adaptations we made to the $\textsc{borg}$ algorithm. In section \ref{Cats}, we describe the tests we performed on our implementation and how we constructed the mock catalogues used in the reconstructions. In section \ref{sec::results}, we display the results of tests. We end with a discussion of our work and our conclusions in section \ref{sec::Conc}. 

\section{Method} \label{sec::method}
This section describes the method for Bayesian physical velocity field reconstruction. Section \ref{PVTracers} discusses a simplification we make with regards to the data, and section \ref{sec::DataModel} describes the model. The physical model used for the velocity field is defined in section \ref{sec::PhysicalModel}, and the prior is specified in section \ref{sec::prior}. The posterior is derived in section \ref{sec::posterior}. We also perform a linear reconstruction as a comparison, which is described in section \ref{WFApproach}. Section \ref{BORG} details the $\textsc{borg}$ architecture.

\subsection{Peculiar Velocity Tracers}
\label{PVTracers}
The data used in velocity field reconstruction are usually galaxies  with a distance indicator: measurements of the distance modulus and of the redshift of a galaxy, from which the comoving distance and galaxy peculiar velocity can be estimated with some associated error.

\textcolor{black}{Such data can be subject to Malmquist bias. The term `Malmquist bias' encompasses several sources of bias, often referred to as selection bias, homogeneous Malmquist bias, and inhomogeneous Malmquist bias. These biases arise respectively due to selection effects in galaxy surveys, underestimates of galaxy distances due to the assumption of local homogeneity, and spurious inflows towards small-scale inhomogeneities in the matter distribution, the latter two due to observational uncertainties. Bias can also be introduced due to the errors on the distance indicators being lognormally-distributed \citep[see e.g.][for further discussion]{1988ApJ...326...19L,1995PhR...261..271S,Sorce2015,Hoffman2021}. Proper handling of Malmquist bias is vital when attempting reconstructions from real data. See e.g. \citet{2016MNRAS.457..172L,2019MNRAS.488.5438G,Boruah2021,2022MNRAS.513.5148V} for examples of reconstructions that handle Malmquist bias.} The effect of Malmquist bias is to systematically shift the inferred peculiar velocity in regions where the model for the galaxy distance distribution is poor, in a way that can be modelled. In regions where the model for the galaxy distance distribution is good, this bias should be negligible, and reconstructed velocities should be a good estimate of the true underlying velocity field. 

\citet{2020MNRAS.498.2703B} compiled the 2nd Amendment type-Ia supernova peculiar velocity catalogue, with tracers up to $z \approx 0.06$. Using this catalogue, they compared a $\chi^{2}$ fit and a full forward-modelling approach that modelled tracer distances, taking into account uncertainties in their reconstructions. Doing so, they found consistent results for $\beta = \frac{f}{b}$ where $b$ is the galaxy linear bias parameter, and $f$ describes the growth rate (see section \ref{sec::DataModel}). Thus, for nearby tracers where distances are well-controlled, we can neglect the uncertainty in galaxy distances without introducing significant biases into reconstructions. 

In this work, we make such an assumption to produce a convenient simplification for the mock tracers. We consider the comoving distance to each tracer to be known exactly, but still consider an error on the peculiar velocity of each tracer. Doing so allows us to build a simple field-based likelihood for the peculiar velocity field that does not require us to deal with uncertainties in distance measurements of tracers at this time. We refer to such idealised tracers as `peculiar velocity tracers' to distinguish them from the usual tracers. \textcolor{black}{As any real distance modulus observation will have a non-zero error, our approach is currently restricted to idealised tracers, but we discuss relaxing the assumption of exactly known distances in future work in section \ref{sec::futurework}.}

\subsection{The Structure Formation Model}
\label{sec::DataModel}

\begin{figure}
  \begin{center}

    \begin{tikzpicture}
    	\pgfdeclarelayer{background}
    	\pgfdeclarelayer{foreground}
    	\pgfsetlayers{background,main,foreground}
    
        \tikzstyle{probability}=[draw, thick, text centered, rounded corners, minimum height=1em, minimum width=1em, fill=darkgreen!20]
    	\tikzstyle{deterministic}=[draw, thick, text centered, minimum height=1.8em, minimum width=1.8em, fill=violet!20]
    	\tikzstyle{variabl}=[ellipse, draw, thick, text centered, minimum height=1em, minimum width=1em]
    	\tikzstyle{data}=[ellipse, draw, thick, text centered, minimum height=1em, minimum width=1em, fill=blue!20]
    
    	\def\blockdist{0.7}
    
        \node (ic) [probability]
        {$\proba\left( \boldsymbol\delta^{\mathrm{I}}|\mathbfss{S} \right)$};
        \path (ic.south)+(0,-\blockdist) node (deltaic) [variabl]
        {$\boldsymbol\delta^{\mathrm{I}}$};
        \path (deltaic.south)+(0,-\blockdist) node (deltalpt) [deterministic]
        {$\boldsymbol\delta^{\mathrm{F}}(\boldsymbol\delta^{\mathrm{I}})$};
        \path (deltalpt.south)+(0,-\blockdist) node (deltalpt_field) [variabl]
        {$\boldsymbol\delta^{\mathrm{F}}$};
        \path (deltalpt_field.south)+(0,-\blockdist) node (vfield_model) [deterministic]
        {$\textbf{v}(\boldsymbol\delta^{\mathrm{F}})$};
        \path (vfield_model.south)+(0,-\blockdist) node (vfield) [variabl]
        {$\textbf{v}$};
        \path (vfield.south)+(0,-\blockdist) node (interp) [deterministic]
        {$V^{\mathrm{r}}(\textbf{v})$};
        \path (interp.east)+(3,0) node (v_pred) [variabl]
        {$V^{\mathrm{r}}$};
         \path (v_pred.north)+(0,\blockdist) node (like) [probability]
        {$\proba\left( V^{\mathrm{PV}} | V^{\mathrm{r}} \right)$};
        \path (like.north)+(0,\blockdist) node (noise) [variabl]
        {$\sigma_{\mathrm{noise}}$};
        \path (noise.north)+(0,\blockdist) node (data) [data]
        {$V^{\mathrm{PV}}$};

    	\path [draw, line width=0.7pt, arrows={-latex}] (ic) -- (deltaic);
    	\path [draw, line width=0.7pt, arrows={-latex}] (deltaic) -- (deltalpt);
    	\path [draw, line width=0.7pt, arrows={-latex}] (deltalpt) -- (deltalpt_field);
    	\path [draw, line width=0.7pt, arrows={-latex}] (deltalpt_field) -- (vfield_model);
    	\path [draw, line width=0.7pt, arrows={-latex}] (vfield_model) -- (vfield);
    	\path [draw, line width=0.7pt, arrows={-latex}] (vfield) -- (interp);
    	\path [draw, line width=0.7pt, arrows={-latex}] (interp) -- (v_pred);
    	\path [draw, line width=0.7pt, arrows={-latex}] (v_pred) -- (like);
    	\path [draw, line width=0.7pt, arrows={-latex}] (noise) -- (like);
    
    	\draw [line width=0.7pt, -latex] (data) to [bend right=70] (like);

    \end{tikzpicture}
  \end{center}
\caption{Hierarchical representation of the Bayesian inference framework used for the analysis of peculiar velocity data with a structure formation model. $\mathbfss{S}$ represents the prior covariance matrix. The  initial density fluctuations, $\boldsymbol\delta^{\mathrm{I}}$, are encoded at $a=10^{-3}$. These are evolved according to a gravity model to form the $z=0$ density contrast, $\boldsymbol\delta^{\mathrm{F}}$. The $z=0$ peculiar velocity field, $\textbf{v}$, is calculated on the grid. This field is interpolated to the positions of tracers, and the line-of-sight component is taken to produce the predicted line-of-sight velocity at tracer positions, $V^\mathrm{r}$. The predicted line-of-sight velocity is compared to mock observations, $V^{\mathrm{PV}}$, assumed to be  Gaussian-distributed with respect to $V^\mathrm{r}$ with a standard deviation $\sigma_{\mathrm{noise}}$. Green boxes denote probability distributions. White ellipses are variables. Purple boxes denote deterministic transitions. Blue ellipses denote observational data.
 \label{fig:BHM}}
\end{figure}
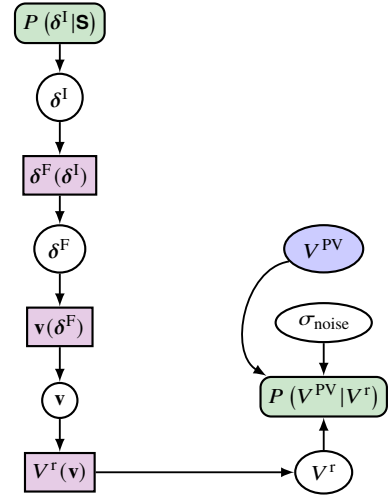

In this section we describe the Bayesian hierarchical model used in this work, which is represented graphically in Figure \ref{fig:BHM}. We refer to this model as the structure formation model.

\subsubsection{The Physical Model} \label{sec::PhysicalModel}
Here we introduce a data model for velocity field reconstructions where the dark matter density contrast evolves to $z=0$ non-linearly using first-order LPT, also known as the Zel'dovich approximation \citep{1970A&A.....5...84Z}. For a comprehensive review of perturbation theory, see \citet{2002PhR...367....1B}. We sketch some details here.

In LPT, the object of interest is the displacement field, $\mathbf{\Psi}(\mathbf{q},t)$, which maps particles from their initial position $\mathbf{q}$ to their final position $\mathbf{x}$ at cosmic time $t$:
\begin{equation}
    \mathbf{x}(t) = \mathbf{q} + \mathbf{\Psi}(\mathbf{q},t) .
\end{equation}
Here, $\mathbf{\Psi}$ satisfies the non-linear equation
\begin{equation}
    J(\textbf{q},t) \nabla_{\mathbf{x}} \cdot \left[\frac{\partial^{2} \mathbf{\Psi}}{\partial t^{2}} + H(t) \frac{\partial \boldsymbol{\Psi}}{\partial t}\right]\ = \ \frac{3}{2} \Omega_{\mathrm{m}}(t) H(t) \left[J(\mathbf{q},t) - 1\right] ,
    \label{LagEqMotion}
\end{equation}
where $J$ denotes the Jacobian of the transformation between  Lagrangian and Eulerian coordinates, $\Omega_{\mathrm{m}}(t)$ is the cosmic matter energy density parameter, and $H(t)$ is the time-dependent Hubble parameter. Requiring that the Lagrangian mass element be conserved between coordinate frames, the Jacobian is
\begin{equation}
    J(\textbf{q},t) = \frac{1}{1 + \delta(\mathbf{x},t)} \ ,
\end{equation}
where $\delta$ is the dark matter density contrast.

The Zel'dovich approximation is the linear solution for the divergence of $\mathbf{\Psi}$, which we denote as $\psi$. Here,
\begin{equation}
    \psi \equiv \nabla_{\mathbf{q}} \cdot \mathbf{\Psi}(\mathbf{q},t) = -D_{1}(t)\delta^{\mathrm{I}}(\mathbf{q}) ,
    \label{ZAsol}
\end{equation}
where $D_{1}$ is the linear growth factor, and $\delta^{\mathrm{I}}(\mathbf{q})$ describes the initial density contrast in Lagrangian coordinates. Therefore, in the Zel'dovich approximation, particles evolve according to 
\begin{equation}
	\textbf{x} = \textbf{q} - D_{1}(t)\nabla^{-1}_{\mathbf{q}} \delta^{\mathrm{I}}(\mathbf{q}) .
	\label{ZAEqMotion}
\end{equation}
As equations \eqref{LagEqMotion}-\eqref{ZAEqMotion} are intrinsically non-linear in the density contrast, the Zel'dovich approximation solution captures some of the non-linearity of the gravitational evolution.

Our model starts at the initial conditions on a regular grid of $N^{3}$ cubic voxels. We denote the value of the initial conditions in the voxel labelled $p$ by $\boldsymbol{\delta}^{\mathrm{I}}_{p}$, where the discretised $\boldsymbol\delta^{\mathrm{I}}$ is a vector of size $N^3$. Next, the grid is populated with dark matter particles, which are subsequently evolved to $z=0$ using the Zel'dovich approximation solution (equation \eqref{ZAEqMotion}). A cloud-in-cell (CiC) deposition scheme \citep{CiC} is employed to produce the $z=0$ density contrast on the grid, where the value in voxel $p$ is denoted by $\delta^{\mathrm{F}}_{p}$.

The velocity field $\textbf{v}$ and the density contrast field $\delta^{\mathrm{F}}$ at a scale factor $a$ can be related by the linearised continuity equation,
\begin{equation}
    \nabla_\textbf{x} \cdot \textbf{v} = - H(a) f(a) \delta^{\mathrm{F}} ,
    \label{EPTvel}
\end{equation}
where $H(a)$ is the Hubble parameter, and $f(a)$ is the growth rate, defined by
\begin{equation}
    f(a) \equiv \frac{\mathrm{d}  \ \mathrm{ln}(D_{1})}{\mathrm{d} \ \mathrm{ln}(a) } .
    \label{lineargrowthfactorlogdev}
\end{equation}
In this work, we evaluate the velocity field at $z=0$.
In first-order Eulerian perturbation theory, the vorticity of the dark matter flow decays as $\frac{1}{a}$, and it has no source term, so we assume that primordial vorticity is negligible, and can therefore define a velocity potential $\theta$ such that
\begin{equation}
    \textbf{v} = -\nabla_\textbf{x} \theta ,
    \label{vpot}
\end{equation}
which satisfies
\begin{equation}
    \nabla^{2}_\textbf{x} \theta = H f \delta^{\mathrm{F}} .
\end{equation}
In our implementation, we use discretised versions of these equations, which involve $\boldsymbol\delta^\mathrm{F}$ (a vector of size $N^3$) and $\textbf{v}$ (a vector of size $3N^3$). Denoting the Discrete Fourier Transform (DFT) of a field $A$ as $\Tilde{A}$, we have
\begin{equation}
\Tilde{\textbf{v}}_{t} = -\mathrm{i}\frac{\textbf{k}_{t}}{|\textbf{k}_{t}|^{2}} H f \Tilde{\boldsymbol{\delta}}^{\mathrm{F}}_{t} \equiv \textbf{G}(\textbf{k}_{t}) H f \Tilde{\boldsymbol{\delta}}^{\mathrm{F}}_{t} ,
    \label{v_field_rec}
\end{equation}
where $t$ is a Fourier grid index, $\textbf{k}_t$ is the associated wavevector, and $\textbf{G}(\textbf{k}) \equiv -\mathrm{i} \textbf{k}/|\textbf{k}|^2$ is the Green's function of the acceleration operator ($\nabla_\textbf{x} \nabla^{-2}_\textbf{x}$). Equation \eqref{v_field_rec} shows that we can linearly relate the Fourier coefficients of the velocity and density contrast fields, allowing us to construct one from the other. Therefore, in configuration space,
\begin{equation}
     \textbf{v}_{p} = H f \sum_{ts} (\mathbfss{F}^{-1})_{pt} \, \textbf{G}(\textbf{k}_{t}) \, \mathbfss{F}_{ts} ~ \boldsymbol{\delta}^{\mathrm{F}}_{s},
    \label{discvfield}
\end{equation}
where $\mathbfss{F}$ is the matrix denoting the DFT operation.

Next, we require the velocity field evaluated at the position of each tracer. We interpolate the velocity field on the grid to each tracer's position using a trilinear interpolation scheme,
\begin{equation}
    \textbf{V}_{n} = \sum_{m} W\left(\textbf{x}_{m} - \textbf{y}_{n}\right) \textbf{v}_{m}
    \label{CIC}
\end{equation}
where $m$ is the grid voxel index, $n$ is a tracer index, $W$ denotes the three-dimensional CiC kernel, ${\textbf{x}}$ and ${\textbf{y}}$ denote the position vectors with respect to the observer for the grid voxels and the tracers respectively. We denote the velocity field at the $n$th tracer position by $\textbf{V}_{n}$ to distinguish it from the field on the grid.
Lastly, we take the line-of-sight component of the velocity field at the tracer positions:
\begin{equation}
    V^{\mathrm{r}}_{n} = \mathbf{V}_{n} \cdot \hat{\mathbf{y}}_{n}
    \label{LOS}
\end{equation}
where $\hat{\textbf{y}} \equiv \textbf{y}/|\textbf{y}|$ indicates a unit vector.

First-order LPT only takes into account some of the non-linearity that we would expect to come about during gravitational collapse. Other structure formation models exist that include additional non-linearity. Accounting for additional non-linearity in the evolution of the density contrast allows for a more physical description of the  smaller-scale behaviour compared to LPT. Furthermore, in general, we need not assume a linear relationship between $\boldsymbol\delta^\mathrm{F}$ and $\textbf{v}$. In section~\ref{sec::futurework}, we discuss possible upgrades of the data model presented in this section.

\subsubsection{The Prior Distribution} \label{sec::prior}
As observations have shown that the primordial density field is well-modelled by a Gaussian random field, $\textsc{borg}$ employs a Gaussian prior for the density contrast at $a=10^{-3}$, which is given by
\begin{equation}
    \proba(\boldsymbol\delta^{\mathrm{I}}|\mathbfss{S}) = \frac{1}{\sqrt{\left| 2 \pi \mathbfss{S} \right|}} \mathrm{exp} \left( -\frac{1}{2} \sum_{pq} \boldsymbol\delta^{\mathrm{I}}_{p} \mathbfss{S}_{pq}^{-1} \boldsymbol\delta^{\mathrm{I}}_{q} \right)\; ,
    \label{BORGprior}
\end{equation}
where {$\mathbfss{S}$} is the prior covariance matrix. Assuming homogeneity, {$\mathbfss{S}$} is a diagonal covariance matrix in Fourier space, and by isotropy, depends only on $|\textbf{k}|$. We assume the fitting formula of \citet{1998ApJ...496..605E,1999ApJ...511....5E} using fiducial values of cosmological parameters (see section \ref{Cats}).

\subsubsection{The Posterior Distribution} \label{sec::posterior}
As described in section \ref{PVTracers}, we use the measured radial velocity of peculiar velocity tracers as the data vector, which we denote $\textbf{V}^{\mathrm{PV}} \equiv \{ V^\mathrm{PV}_n \}_{1 \leq n \leq N_\mathrm{tracers}}$ for a set of $N_\mathrm{tracers}$ tracers. Using the primordial density prior given by equation \eqref{BORGprior}, we write the posterior using Bayes' theorem,
\begin{equation}
    \proba\left( \boldsymbol\delta^{\mathrm{I}} | \textbf{V}^{\mathrm{PV}} \right) = \frac{\proba\left( \textbf{V}^{\mathrm{PV}} | \boldsymbol\delta^{\mathrm{I}}\right) \proba\left(\boldsymbol\delta^{\mathrm{I}}\right)}{\proba\left( \textbf{V}^{\mathrm{PV}}\right)} .
    \label{Bayes}
\end{equation}
To write down the likelihood, we make use of the fact that $\textsc{borg}$ uses a deterministic physical model. This model allows us to connect the initial conditions to the line-of-sight velocity field evaluated at the position of each tracer. Assuming that the peculiar velocity data only depend on the initial conditions through the line-of-sight component of the final velocity field evaluated at tracer positions $\textbf{V}^\mathrm{r} \equiv \{ V^\mathrm{r}_n \}_{1 \leq n \leq N_\mathrm{tracers}}$, we can expand the likelihood as
\begin{equation}
    \begin{split}
    \proba \left( \textbf{V}^{\mathrm{PV}} | \boldsymbol\delta^{\mathrm{I}} \right) 
    & = \int \proba\left( \textbf{V}^{\mathrm{PV}}, \textbf{V}^\mathrm{r} | \boldsymbol\delta^{\mathrm{I}} \right) \mathrm{d} \textbf{V}^\mathrm{r}     \\
   & = \int \proba\left( \textbf{V}^{\mathrm{PV}}| \textbf{V}^\mathrm{r} \right) \proba\left( \textbf{V}^\mathrm{r} | \boldsymbol\delta^{\mathrm{I}} \right) \mathrm{d} \textbf{V}^\mathrm{r}   .
    \end{split}
\end{equation}
We denote the operator that produces the line-of-sight velocity field at the $n$th tracer position from the initial conditions by $V^{\mathrm{r}}_{ n}\left(\boldsymbol\delta^{\mathrm{I}} \right)$. Its form is described in section \ref{sec::PhysicalModel}. As the physical model is deterministic, we can write 
\begin{equation}
    \proba\left( \textbf{V}^\mathrm{r} | \boldsymbol\delta^{\mathrm{I}} \right) = \prod_{n=1}^{N_{\mathrm{tracers}}} \delta_\mathrm{D} \left[ V^{\mathrm{r}}_{n} - V^{\mathrm{r}}_{n}\left(\boldsymbol\delta^{\mathrm{I}} \right) \right]
\end{equation}
where $\delta_\mathrm{D}$ is the Dirac delta distribution. Assuming the noise for each tracer is independent,
\begin{equation}
\proba \left( \textbf{V}^{\mathrm{PV}} | \boldsymbol\delta^{\mathrm{I}} \right) = \prod_{n=1}^{N_\mathrm{tracers}} \proba\left[ V^\mathrm{PV}_n | V^{\mathrm{r}}_{n}\left(\boldsymbol\delta^{\mathrm{I}} \right) \right] .
\end{equation}
Further, we assume independent Gaussian likelihoods for each tracer, giving
\begin{equation}
    \proba \left( \textbf{V}^{\mathrm{PV}} | \boldsymbol\delta^{\mathrm{I}} \right) = \prod_{n=1}^{N_{\mathrm{tracers}}} \frac{1}{\sqrt{2 \pi \sigma^2_{n}}} \exp\left\lbrace-\frac{1}{2} \frac{\left[ V^\mathrm{PV}_n - V^{\mathrm{r}}_{n}\left( \boldsymbol\delta^{\mathrm{I}} \right)  \right]^{2}}{\sigma^{2}_{n}} \right\rbrace.
    \label{VIRBIUS_likelihood}
\end{equation}
Here,  $\sigma_{n}$ is the uncertainty in the measured line-of-sight velocity of the $n$th tracer. In general, $\sigma_{n}$ would have contributions from the uncertainty in the measured tracer redshift, the measured tracer distance modulus, and any theoretical uncertainty arising from non-linearity that the model fails to take in to account. Using the simplified notion of peculiar velocity tracers, the uncertainty here instead accounts for the uncertainty in the peculiar velocity, as well as any unaccounted-for non-linearity. We assume a single value of $\sigma_n$ for all tracers, which we denote $\sigma$.

\subsection{The Linear Model} \label{WFApproach}
Alongside the structure formation model, we perform reconstructions using a linear data model and the Wiener filtering approach for comparison. See \citet{1995ApJ...449..446Z} for a more in-depth discussion of the method. In order to reduce confusion, we refer to this Wiener filtering approach as the linear implementation. 

First, we make a distinction. The problem of Wiener filtering can be solved by two different approaches. The first approach involves optimising for the maximum a posteriori solution. The second involves sampling the Wiener posterior and estimating the sample mean and variance. In this work, we use the latter approach so as to compare to the structure formation model case.

A linear data model requires the prediction to be linearly related to the target variables of the inference:
\begin{equation}
	\textbf{V}^{\mathrm{r}} = \mathbfss{R} \boldsymbol\delta^{\mathrm{I}},
\end{equation}
where $\mathbfss{R}$ is a linear operator. In order to write down $\mathbfss{R}$, we require the initial dark matter density contrast $\boldsymbol\delta^{\mathrm{I}}$ to evolve linearly with cosmic time. In this case, the values of the initial and final density contrast in the $p$th voxel are related by
\begin{equation}
    \boldsymbol\delta^{\mathrm{F}}_{p} \ = \ D_{1}(t_0) \boldsymbol\delta^{\mathrm{I}}_{p} .
    \label{EPTden}
\end{equation}
Then, as in the structure formation data model, we construct the velocity field on the grid using the linear operator of equation \eqref{discvfield}. We use a CiC interpolation to calculate the velocity at the position of each tracer as described in equation \eqref{CIC}. Finally, we take the line-of-sight component of the velocity at the position of each tracer, as given by equation \eqref{LOS}. This sequence of linear operations defines the $\mathbfss{R}$ operator.

For the likelihood, the Wiener filtering approach requires us to use a Gaussian likelihood. We use a likelihood of the same functional form as the one used in the structure formation case, given by equation \eqref{VIRBIUS_likelihood}. We also require a Gaussian prior, given by equation \eqref{BORGprior}. As a result, the posterior will be of the same functional form as in the structure formation case: any differences will be due to the different forms of the operators $V^{\mathrm{r}}_n\left(\boldsymbol\delta^{\mathrm{I}} \right)$.

\subsection{\textsc{borg}} \label{BORG}
In its original form, the $\textsc{borg}$ algorithm infers the initial conditions constrained by galaxy number counts. More detailed descriptions of the $\textsc{borg}$ framework and its applications can be found in \citet{2013MNRAS.432..894J,2015JCAP...01..036J,2016MNRAS.455.3169L,2019A&A...625A..64J,2019arXiv190906396L}.

High-resolution inference can involve $\mathcal{O}$(10$^7$) or more parameters, and to sample such high-dimensional posteriors, $\textsc{borg}$ makes use of Hamiltonian Monte Carlo (HMC) sampling \citep{1987PhLB..195..216D,Neal1993,Neal1996}. For more information on the $\textsc{borg}$ HMC sampler, see \citet{2010MNRAS.407...29J,2013MNRAS.432..894J}. Different implementations of HMC have previously been used in the context of velocity field reconstruction \citep{Boruah2021,2022MNRAS.513.5148V}. For the purpose of this work, we adapted the existing machinery of $\textsc{borg}$ to infer the initial density constrained by peculiar velocity observations. We used the $\textsc{borg}$ HMC framework to sample the posterior given by equation \eqref{Bayes}. HMC sampling requires knowledge of the gradient of the posterior with respect to the inferred parameters, which we derive in Appendix~\ref{AdjGrad}.

We also implemented the linear data model within the $\textsc{borg}$ code, and we used the $\textsc{borg}$ HMC sampler to sample the corresponding posterior. We derive the gradient of the linear model in Appendix \ref{AdjGrad}.

As a post-processing step, re-applying the structure formation model to each sample of the initial conditions $\boldsymbol\delta^{\mathrm{I}}$ produces samples of the $z=0$ density contrast $\boldsymbol\delta^{\mathrm{F}}$ and velocity field $\textbf{v}$. The ensemble of such samples provides the full posterior predictive distributions for $\boldsymbol\delta^{\mathrm{F}}$ and $\textbf{v}$, which can be summarised by field means and variances. 

\subsection{Model Mis-specification} \label{sec::MM}

\textcolor{black}{
An ever-present issue in performing reconstructions is that of model mis-specification, which arises when the model used in the data analysis differs from the actual data-generating process. This is a problem present in all inference-based cosmology, as the quality of any results learned from a dataset is limited by the accuracy of the model used to analyse it. As all physical models are inherently approximations, every inference on real data is in some sense limited by model mis-specification. In the context of structure formation, the model is mis-specified if, for example, we use mock data generated with a $N$-body simulator, but perform the inference with an LPT structure formation model.} \citet{2021JCAP...03..058N} have previously investigated the impacts of various types of model mis-specification on forward inferences of the initial conditions. 

\textcolor{black}{
In this work, we aim to address the problem of model mis-specification in reconstructions of the initial conditions constrained by peculiar velocity tracers, and test the impacts of mis-specification on inferred constraints of the initial conditions. We choose to account for model mis-specification in a way similar to $\textsc{virbius}$, by adding a Gaussian scatter in quadrature to the modelled scatter due to noise, to give the total uncertainty in the likelihood.
\citet{2016MNRAS.457..172L,2019MNRAS.488.5438G,2022MNRAS.513.5148V} make use of a linear physical model which does not capture non-linear effects in the data. Therefore, they refer to their additional Gaussian scatter as $\sigma_{\mathrm{NL}}$, where `NL' stands for `Non-Linearity'. Our LPT model is mildly non-linear, and so intrinsically captures some of the non-linearity in the data. Thus, we denote our additional Gaussian scatter by $\sigma_{\mathrm{MS}}$, where `MS' stands for `mis-specification', to highlight that in our approach, some of the description of the non-linear regime is present in the data model rather than absorbed by the likelihood.}

Denoting the scatter due to noise as $\sigma_{\mathrm{noise}}$, and the total likelihood uncertainty as $\sigma$, we have
\begin{equation}
    \sigma = \sqrt{ \sigma_{\mathrm{noise}}^{2} + \sigma_{\mathrm{MS}}^{2}}.
\end{equation}
\textcolor{black}{
Since the addition of an additional random noise is evidently not a perfect solution for the consequences of using a simplified gravity model, in section \ref{sec::futurework} we discuss further possible improvements to our physical model to allow us to better describe non-linear structure formation within the inference, and so potentially reduce the value of $\sigma_{\mathrm{MS}}$ needed.}

\section{Mock Catalogues and Simulations} \label{Cats}

In order to test the structure formation implementation, we created a number of mock peculiar velocity tracer catalogues. All tracers in each catalogue are located in a cube of side length $L = 250\Mpch$. For each catalogue, we assumed a flat $\Lambda$CDM cosmology and generated the initial density fluctuations using a cosmological matter power spectrum described by \citet{1998ApJ...496..605E,1999ApJ...511....5E}.
The cosmological parameters are fixed for all catalogues. We used the Planck 2018 best-fitting values \citep{2020A&A...641A...6P}, given in Table \ref{tab:cpar}.

The first set of catalogues consists of peculiar velocity tracers drawn from a single velocity field realisation at $z=0$, made using the structure formation data model. This set serves as a self-consistency check for the structure formation method. Here, we generated a realisation of the initial conditions on a grid of $128^{3}$ voxels, as described above. \textcolor{black}{The initial conditions were populated with $256^3$ dark matter particles, which were evolved to $z=0$ using LPT. The $z=0$ density constrast was then constructed from the particles using CiC binning. Using equation \eqref{discvfield}, the $z=0$ velocity field on the grid was constructed from the $z=0$ density contrast.} Choosing to draw tracer positions uniformly within the box, we used a CiC interpolation to find the velocity field at each tracer position, and computed the line-of-sight component. The catalogues contained 500 and 10000 tracers each. 
\textcolor{black}{Finally, the mock data were obtained by adding to each tracer's radial peculiar velocity noise drawn from a Gaussian distribution with zero mean and standard deviation $\sigma_\mathrm{noise} = 150~\kms$ \citep[as in][]{2015MNRAS.450..317C}. Prior to the addition of the simulated noise, these data are made exactly with the model described in section \ref{sec::DataModel}, and thus there is no model mis-specification for inference with the LPT model.  Therefore in section \ref{sec::scresults}, we use $\sigma_\mathrm{MS}=0$ (i.e. $\sigma = \sigma_\mathrm{noise}$) in the likelihood.}

The second catalogue consists of peculiar velocity tracers drawn from a mock halo catalogue. \textcolor{black}{This catalogue is used to test whether broadening the likelihood by including an additional Gaussian scatter can account for the model mis-specification due to the extra non-linearity.} \textcolor{black}{Starting from a set of initial conditions on a grid of $128^{3}$ voxels, the initial conditions were populated with $512^{3}$ dark matter particles, which were evolved to $z=64$ with LPT} using the cosmological simulation package $\textsc{simbelmyn{\"e}}$ \citep{2015JCAP...06..015L}. We fed the particles at $z=64$ to the $N$-body simulation code $\textsc{gadget2}$ \citep{2005MNRAS.364.1105S}, which ran to $z=0$ using again a particle-mesh grid of $128^{3}$ points. The friends-of-friends halo finder package $\textsc{rockstar}$ \citep{2013ApJ...762..109B} was used on the resulting $\textsc{gadget2}$ snapshot. We \textcolor{black}{uniformly} selected 10000 of the resulting haloes to make up the halo catalogue. \textcolor{black}{To obtain the mock data, we added Gaussian noise with zero mean and standard deviation $\sigma_\mathrm{noise} = 150~\mathrm{km}\;\mathrm{s}^{-1}$ to each halo's radial peculiar velocity, as in our tests of self-consistency.}

As $\textsc{gadget2}$ is a full $N$-body simulator, the resulting $z=0$ fields encapsulate the non-linear gravitational evolution of the cosmic matter more realistically than using LPT. Furthermore, halo finding is a non-linear operation. 
\textcolor{black}{For these two reasons, an inference using the LPT model and our second catalogue as the data will by subject to model misspecification. Therefore, in section \ref{MM}, we broaden the likelihood by using a non-zero value of $\sigma_\mathrm{MS}$, added in quadrature to $\sigma_{\mathrm{noise}}$ to give the total $\sigma$ used in the likelihood. We discuss the values of $\sigma_{\mathrm{MS}}$ that we investigated in section \ref{sec::hfresults}.}

\begin{table}
\caption{Planck Cosmological parameter values used in the reconstructions.}
\label{tab:cpar}
\begin{tabular}{c|c|c|c|c|c}
\hline
$\Omega_{\mathrm{m}}$ & $\Omega_{\Lambda}$ & $\Omega_{\mathrm{b}}$ & $\sigma_{8}$ & $n_{\mathrm{s}}$ & $h$ \\
\hline 0.315 & 0.685 & 0.0492 & 0.811 & 0.965 & 0.674 \\
\hline
\end{tabular}
\end{table}

\section{Results} \label{sec::results}
In this section, we display the results of field-based inference from the peculiar velocity tracer catalogues described in section \ref{Cats}. Section \ref{sec::scresults} shows the results of the self-consistency tests. Section \ref{MM} presents the results from the halo mock catalogue, where we investigate the robustness of the inference to model mis-specification. Figures produced using samples from chains run using the structure formation model are labelled as ‘$\mathrm{LPT}$', and those produced using samples from chains run using the linear model as ‘$\mathrm{Linear}$’. 

\subsection{Self-Consistency Tests} \label{sec::scresults}

\subsubsection{Sampler Warm-Up} \label{SCburnin}
In order to sample the posteriors, we used the $\textsc{borg}$ HMC sampler. The warm-up behaviour of this sampler is investigated in Appendix~\ref{ap::scburnin}. We find that the sampler decorrelates from the starting point and starts to explore the high-density region of the posterior after roughly 1000 samples when using samples from either the structure formation or linear models applied to the catalogue of 500 or 10000 tracers. The autocorrelation length for the value of the initial conditions within a typical voxel is roughly 200 samples.

\subsubsection{Reconstructed Fields}
\begin{figure*}
\includegraphics[width=0.99\textwidth]{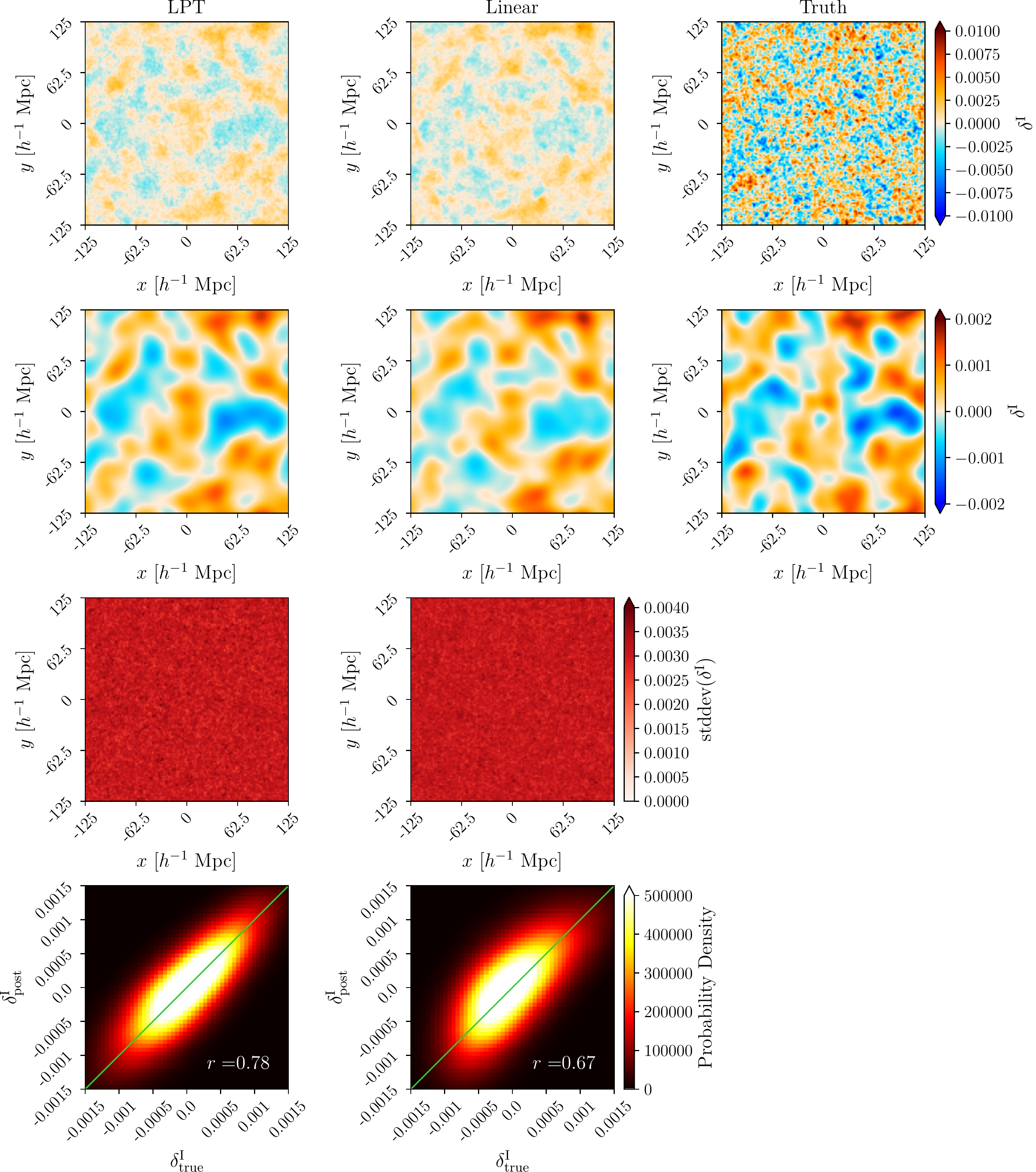}
\caption{\label{SCicslices} Top Row: from left to right, slices of the posterior mean initial conditions for the structure formation model (LPT) and linear model (Linear) with 10000 tracers , and finally the corresponding slice in the field from which the data are drawn (Truth), which we refer to as the `true' field. All are plotted with the same colour scaling. Second Row: the mean fields and true field plotted in the top row smoothed with a Gaussian filter at an $8\Mpch$ scale. Third Row: from left to right, slices of the posterior initial conditions standard deviation for the structure formation model and linear model, plotted with the same colour scaling.  Bottom Row: from left to right, normalised two-dimensional histograms of the initial conditions from samples against the true initial conditions for the structure formation model chain and the linear model chain. We smooth each sample using the Gaussian filter described above and construct a histogram, and average the histograms along the chain. When smoothed, both models visually reconstruct the true initial conditions here with a similar variance. As both histograms align with the green line, the posterior distributions are consistent with the distribution of the true initial conditions. The structure formation model posterior has a greater correlation coefficient than the linear model posterior.
}
\end{figure*}

\begin{figure*}
\centering
\includegraphics[width=0.99\textwidth]{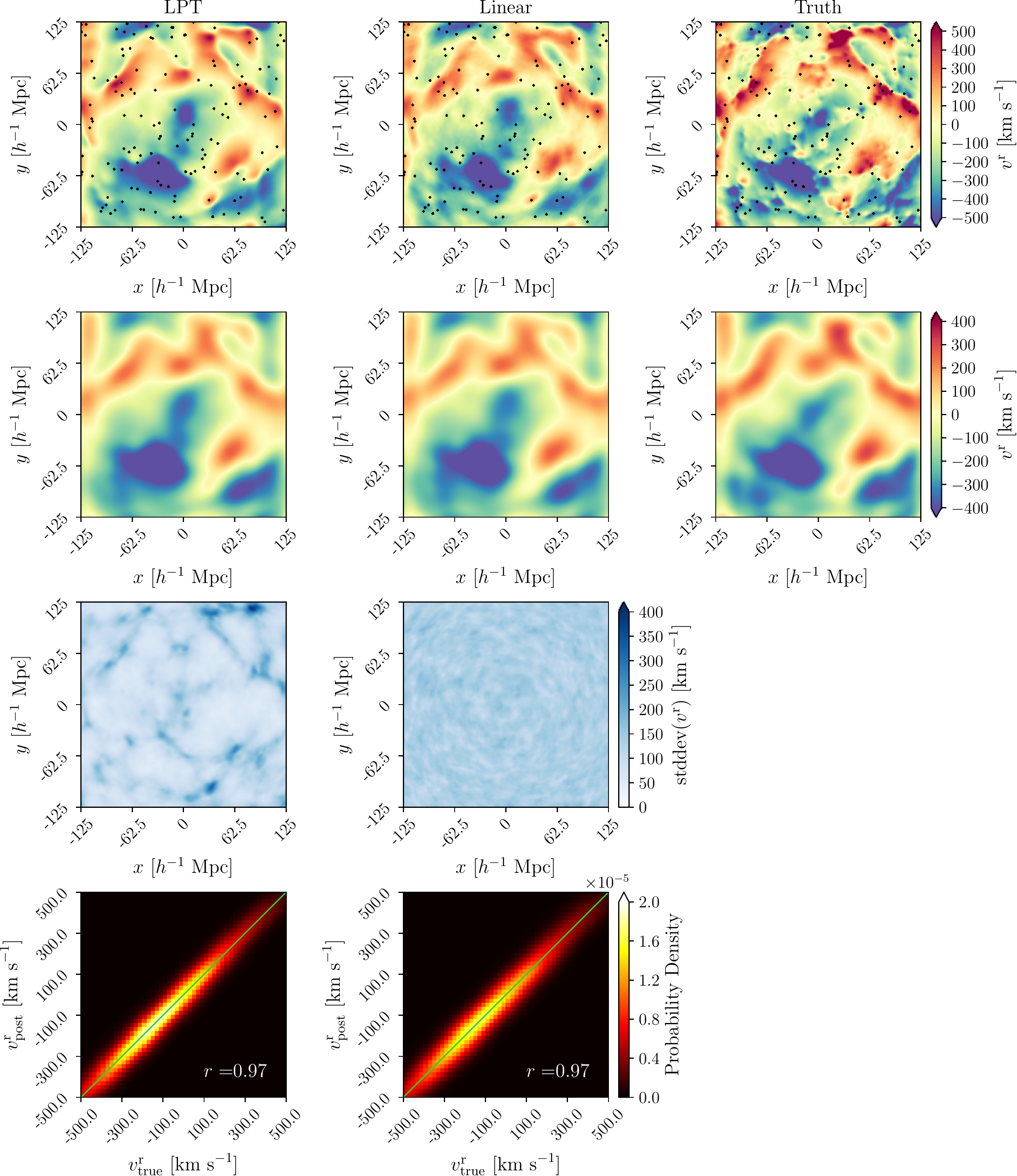}
\caption{\label{SClosslices} Same as Fig. \ref{SCicslices}, but for the posterior present-day line-of-sight velocity field. The tracers with a perpendicular distance less than $1.9 \Mpch$ from this slice are plotted as black dots. After smoothing, both models visually reconstruct the true line-of-sight velocity field. The structure formation model does so with a generally lower variance. Both histograms align with the green line, meaning that both distributions are consistent with the distribution of the true field. However, the structure formation distribution is slightly narrower, implying a smaller scatter. }
\end{figure*}

\begin{figure*}
\centering
\includegraphics[width=0.99\textwidth]{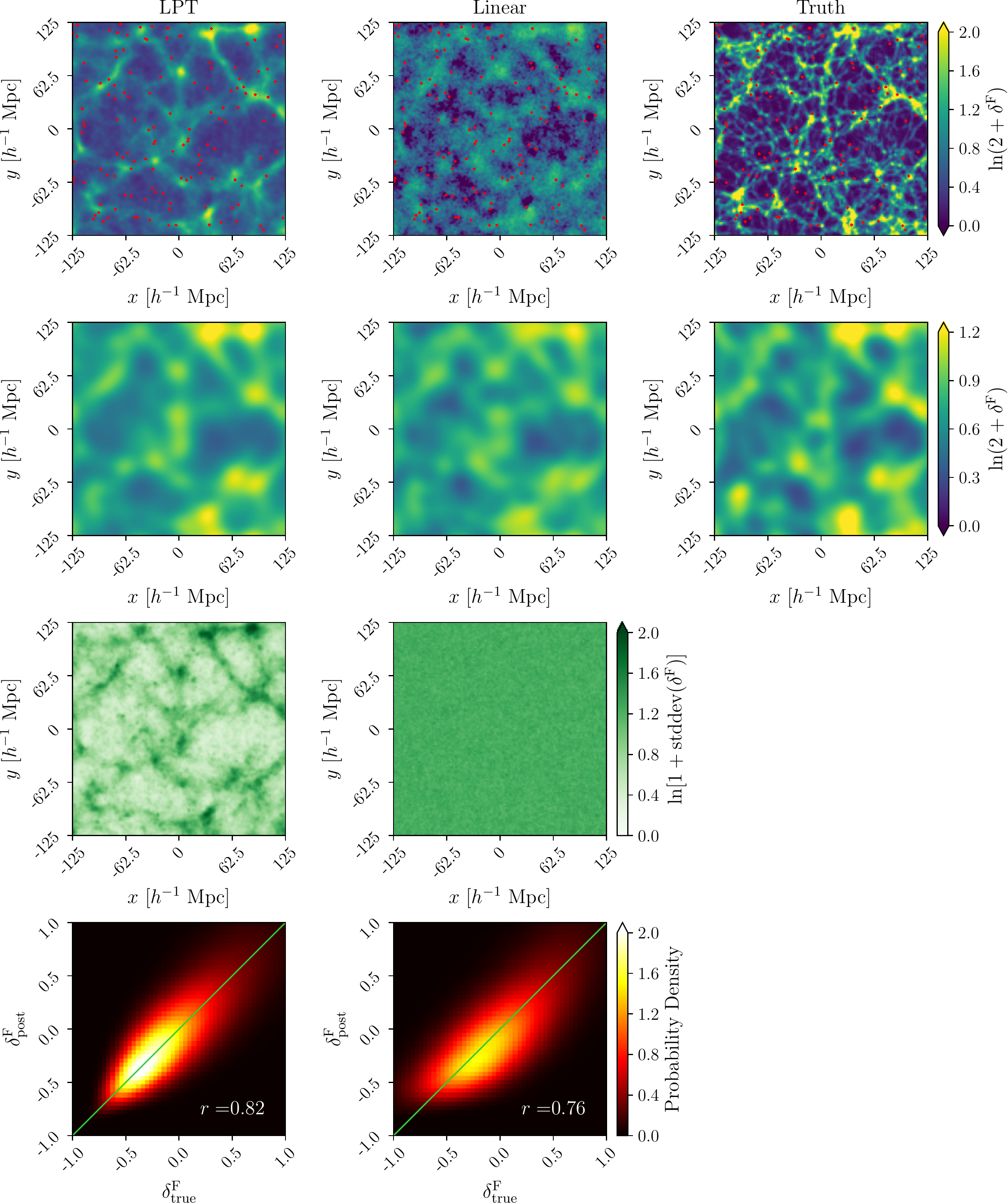}
\caption{\label{SCdenslices} Same as Fig. \ref{SCicslices}, but for the posterior $z=0$ density contrast. The tracers with a perpendicular distance less than $1.9 \Mpch$ from this slice are plotted as red dots. The linear model misses small-scale over-dense structure in the true field. After smoothing, both models visually reproduce the smoothed truth, showing that the linear model only captures the large-scale information. Furthermore, the structure formation model reconstruction has a lower variance across the slice. The histogram for the structure formation model aligns with the green line, meaning that it is consistent with the distribution of the true field. The linear model histogram is slightly misaligned, showing that this model does not fully capture the true distribution down to an $8\Mpch$ scale. }
\end{figure*}

We now qualitatively investigate the reconstructed field samples using the catalogue of 10000 tracers. For both the structure formation and linear models, the inference yields samples of the initial conditions. For each sample, we obtain the corresponding realisation of the final density contrast and line-of-sight velocity field via a forward application of the data model, as discussed in section \ref{BORG}.

In Figures \ref{SCicslices}, \ref{SClosslices}, and \ref{SCdenslices}, on the top row, we show the central slice of the chain mean field for the structure formation model on the left, and for the linear model in the middle. The true field used in the data generation process is on the right. In the second row, we plot the mean and true fields from the top row, after they have been smoothed with a Gaussian filter with a width of $8h^{-1}~\mathrm{Mpc}$, chosen as this is roughly the scale at which the cross-correlation between the chain mean initial conditions and the true initial conditions falls to 0.5 (see Figure \ref{SCspec}, top right panel). The third row shows the standard deviation of the chain in the same slice for the structure formation model on the left, and for the linear model on the right. The bottom row displays normalised two-dimensional histograms of the field from samples against the true field for the structure formation model chain on the left, and for the linear model chain on the right. We smooth each sample using the Gaussian filter described above and construct a histogram, and average the histograms along the chain. The correlation coefficient between the model posterior and true distribution is given in each case.

The reconstructed initial conditions are shown in Figure~\ref{SCicslices}. After smoothing, both models visually resemble the true initial conditions. From the third row, we see that the standard deviation in the slice is comparable in both models. In both cases, the histograms in the bottom row align with the green line, which shows that both data models have posterior distributions consistent with the true distribution at scales greater than $8\Mpch$. The structure formation model posterior has a greater correlation coefficient than the linear model posterior.

Next, the posterior sample mean of the $z=0$ line-of-sight velocity field is shown in Figure~\ref{SClosslices}. Visually, both \textcolor{black}{smoothed} mean fields qualitatively resemble the true field. The standard deviation slices show that the structure formation model has a generally lower variance across the entire slice, and so provides better constraints on the velocity field than the linear model. Both histograms align with the green line, again showing that both posteriors are consistent with the true distribution. However, the distribution is slightly narrower for the structure formation model than the linear model, implying a smaller scatter. 

Finally, the reconstructed $z=0$ density contrast is shown in Figure~\ref{SCdenslices}. The structure formation model mean field visually reproduces a smooth version of the truth, and is able to reproduce small-scale over-dense structures. Unlike the linear model, the structure formation model mean field has a physical value ($\delta^{\mathrm{F}} \geq -1$) in every voxel.

Following smoothing, both models visually reproduce the truth, showing that the linear model is only able to capture the large-scale information here. Again, the structure formation model generally has a lower variance across the entire slice. The histogram for the structure formation model is aligned with the green line, showing that the corresponding posterior is consistent with the true posterior. This is not true for the linear model posterior, which is a shortcoming in the previous state-of-the-art method that our approach involving a structure formation model is able to overcome. The interpretation of this shortcoming is that, in the case of a linear model linking the initial conditions to the velocity field, the final density contrast is (implicitly) Gaussian-distributed, whereas the true field is drawn from a fundamentally different distribution. 

\textcolor{black}{In this work, we use $\sigma_{8} = 0.811$, meaning that full non-linear behaviour should occur for scales smaller than $ \sim 8 \Mpch$. The unsmoothed LPT density contrast shows filamentary structures occurring on scales of $ \sim 10 \Mpch$ to $ \sim 50 \Mpch$. These mildly non-linear structures are an effect of non-linear gravity evolution, and do not appear from Gaussian initial conditions with Eulerian linear evolution. Therefore, their presence shows that non-linearity is present in our LPT reconstructions. }

As expected, the posterior standard deviation for the structure formation model line-of-sight velocity is lower than $150~\mathrm{km}\;\mathrm{s}^{-1}$, the magnitude of the noise, for nearly every voxel in the box. The precision of the reconstructed velocity field is not limited to the precision to which the tracer peculiar velocities are known. This improvement in the average standard deviation of the line-of-sight velocity field, comparing the structure formation model to the linear model, occurs because we make use of a physics-informed likelihood. \textcolor{black}{Using the LPT model, the inference is better able to capture the large-scale correlations present in the data,} 
which helps constraining the velocity field even where the signal is low.

\begin{figure*}
\centering
\includegraphics[width=0.99\textwidth]{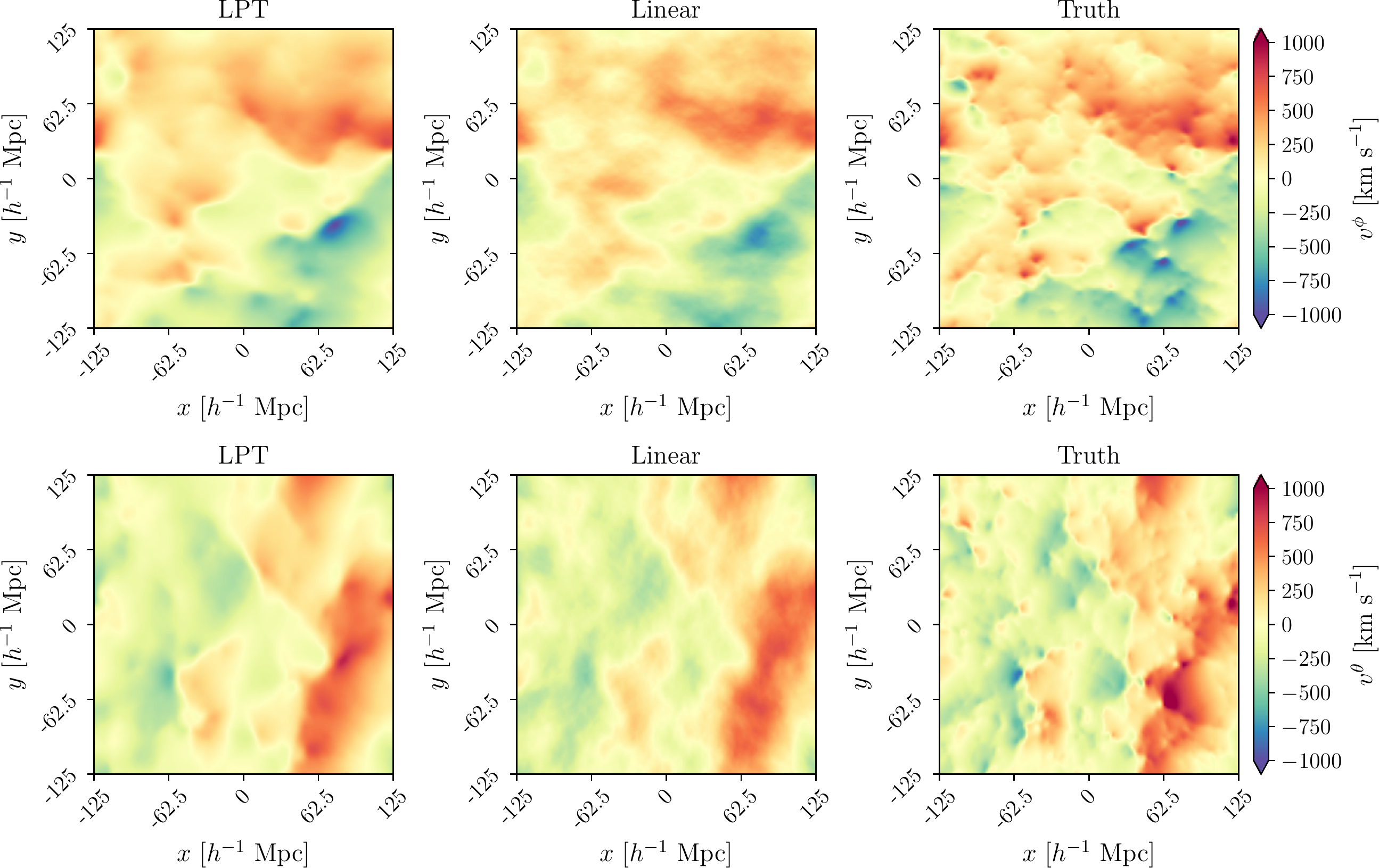}
\caption{\label{SCangslices} \textcolor{black}{Transverse components of the $z=0$ velocity field (in spherical coordinates). As in Fig. \ref{SClosslices}, the left column is the posterior mean for the LPT model, the middle column is the posterior mean for the linear model, and the right column is the ground truth. The azimuthal angle ($\phi$) component is plotted on the top row, and the polar angle ($\theta$) component on the bottom row. Qualitatively, both models reproduce both angular components here, such that, together with Fig. \ref{SClosslices}, each model can qualitatively reproduce the full 3D velocity ground truth.  }}
\end{figure*}

\textcolor{black}{Making use of a non-linear data model means that the inferred signal and the standard deviation can be correlated.}
\textcolor{black}{In the context of our results, we consistently observe that the standard deviation  from the LPT model contains structure, and spatially correlates with the reconstructed signal (see Figure \ref{SCdenslices} left-hand column, first and third rows, for example). This is unlike the Wiener filtering case, where the standard deviation is smooth and unstructured. }
\textcolor{black}{This correlation between signal and standard deviation when using the LPT model is in agreement with previous work: \citet{2015JCAP...01..036J} demonstrate the same effect originating from a non-linear data model in the case of the inference of the initial conditions constrained by SDSS galaxy clustering data (see Figure 6 there, and the associated discussion). Furthermore, from a physical point of view, \citet[Figure~2]{2019MNRAS.487..228B}} demonstrated that the trace of the cosmic velocity dispersion tensor correlates well with the matter overdensity field. Multi-streaming regions, in which the velocity field standard deviation is generally larger, need to be described by models that go beyond the linearised continuity equation (equation \eqref{EPTvel}). We expect that using such a model would reduce the variance in the $z=0$ density contrast and line-of-sight velocity at the positions of structures, allowing for better constraints of each field here. \textcolor{black}{For completeness, in Figure \ref{SCangslices} we also plot the (unsmoothed) transverse components of the posterior mean $z=0$ velocity field, to highlight the fact that the full 3D velocity field is accessible as a latent variable in our Bayesian hierarchical model, not just its radial component.} \textcolor{black}{
Plots of the chain posterior standard deviation of the transverse components of the velocity field show results in keeping with the results for the radial component. Thus, we omit them here.}

\subsubsection{Power Spectra and Cross-Correlation Coefficient} \label{sec::quantresults}
\begin{figure*}
\centering
\includegraphics[width=0.99\textwidth]{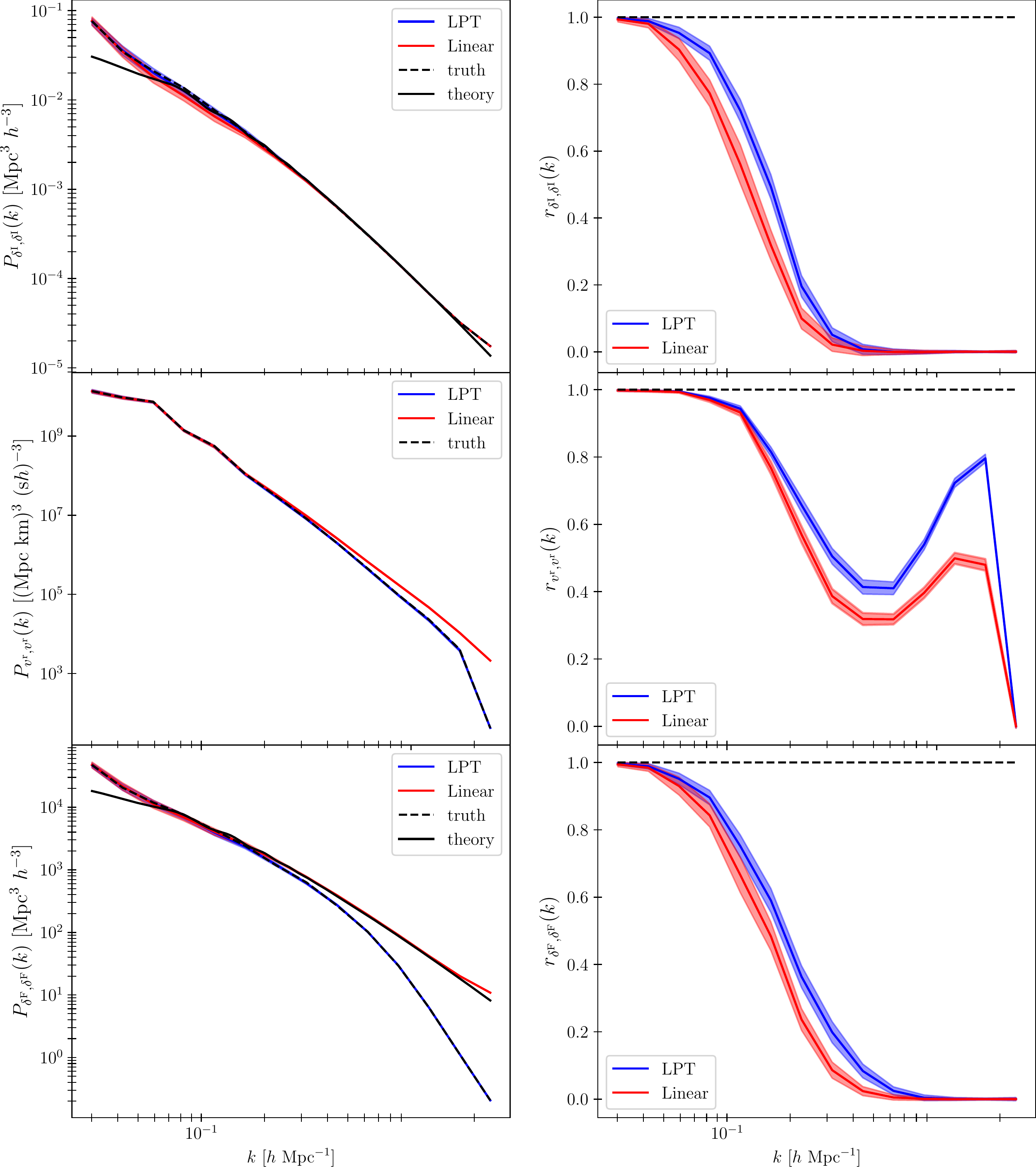}
\caption{\label{SCspec} 
Power spectra of field samples (left column) and the cross-correlation coefficient between samples and the corresponding true field (right column) from chains run on the catalogue of 10000 tracers. The first, second and third rows correspond to the initial density contrast $\delta^\mathrm{I}$, the $z=0$ line-of-sight velocity $v^\mathrm{r}$ and the $z=0$ density contrast $\delta^\mathrm{F}$, respectively.
We use both a model accounting for structure formation (`LPT', in blue), and a linear model (`Linear', in red) in the inference. 
\textcolor{black}{Note that samples were not smoothed prior to the calculation of the power spectra and cross-correlations.}
 In each case, we plot the chain mean and the $2\sigma$ credible region, estimated from samples after warm-up. The dashed black line represents the true fields. For $\delta^\mathrm{I}$, the solid black line represents the theory prediction, and for $\delta^\mathrm{F}$, the solid black line represents the $\delta^\mathrm{I}$ theory line linearly rescaled to $z=0$. Both models can reproduce the power spectra of the true initial conditions. The structure formation model can self-consistently reproduce the statistics of the $z=0$ line-of-sight velocity field and density contrast, whereas the linear model power spectra for the same fields are biased high for $k \gtrsim 0.1h$ Mpc$^{-1}$. The structure formation model cross-correlations are greater than or equal to the linear model cross-correlations for all fields at all scales, showing that the structure formation model reconstructions have at least as good phase accuracy as the linear model. 
}
\end{figure*}

In this section, we quantitatively investigate the two-point statistics of reconstructed fields (initial conditions, final density and final line-of-sight velocity), produced using the catalogue of 10000 tracers. 
\textcolor{black}{The comoving size of the reconstruction box is $L = 250 \Mpch$, which yields a fundamental mode with wavenumber $k = 2 \pi / L = 0.0025 \hMpc$. }
The left-hand column of Figure \ref{SCspec} displays the power spectra of these fields. For each field, we calculate the mean and standard deviation of the power spectra from the samples, and compare each case to the corresponding `true’ field used in the data generation process. Note that in the top-left and bottom-left panels of Figure \ref{SCspec}, the power spectrum of the true field does not align with the Eisenstein \& Hu spectrum at large scales \textcolor{black}{($k \lesssim 0.07 \hMpc$)} due to sample variance. On small scales \textcolor{black}{($k \gtrsim 1 \hMpc$)}, deviations from this theory line are gridding effects. The power spectra of the reconstructed initial conditions are consistent at $2\sigma$ with the power spectrum of the true initial conditions at all available scales, for both the structure formation and the linear model. This means that our structure formation model is able to self-consistently infer the statistics of the initial conditions. The power spectrum of the $z=0$ line-of-sight velocity field gives a quantitative measure of the quality of reconstructed velocity fields. The structure formation model produces fields with power spectra consistent with the power spectrum of the truth at all available scales. On the contrary, the linear model produces power spectra that are too high at small scales \textcolor{black}{($k \gtrsim 0.1h$~Mpc$^{-1}$)}. The same is true for the $z=0$ density contrast. Therefore, the structure formation model self-consistently reconstructs the statistics of the latent variables of the inference problem, unlike the linear model. 

In the right-hand column of Figure \ref{SCspec}, we plot the cross-correlation coefficient between the samples and the corresponding true field. The Fourier-space cross-correlation coefficient is defined by
\begin{equation}
    r(k) \equiv \frac{\langle A_{\mathrm{post}}(k)A^{*}_{\mathrm{true}}(k)\rangle_{k \in k~ \mathrm{bin}}}{\sqrt{P_{A_{\mathrm{post}}}(k) P_{A_{\mathrm{true}}}(k)}}
    \label{cross-corr}
\end{equation}
where $A \in \{\delta^{\mathrm{I}}, \delta^{\mathrm{F}},v^{\mathrm{r}}\}$. $P_{A, \mathrm{post}}$ is the power spectrum of a posterior sample field $A_{\mathrm{post}}$, and $P_{A, \mathrm{true}}$ is the power spectrum of the true field $A_{\mathrm{true}}$. The cross-correlation coefficient is a measure of the phase accuracy of reconstructions. As in the left-hand column, we calculate the mean and standard deviation from each chain.

\textcolor{black}{The cross-correlations between the fields inferred with the LPT model and the truth are clearly better than the correlations between the fields inferred with the Linear model and the truth. The LPT cross-correlations are greater than or equal to the Linear cross-correlations at all scales in the reconstruction box, for all three fields. Between $0.1 \hMpc \lesssim k \lesssim 0.4 \hMpc$, the LPT cross-correlations with the truth is notably higher than the Linear cross-correlations. Therefore, while Figure \ref{SClosslices} shows that the recovered Linear and LPT velocity fields are qualitatively similar, the LPT model quantitatively reproduces the true fields with a generally greater phase accuracy than the Linear model.}

\textcolor{black}{The posterior power spectra for the transverse components of the velocity field show behaviour in keeping with the posterior power spectra of the radial velocity field. We therefore omit them.}

Our inference framework using a structure formation model is able to correctly reproduce the two-point statistics of the initial conditions, $z=0$ density field, and line-of-sight velocity field, as well as produce fields that are qualitatively similar to the true fields. Therefore, we conclude that it is self-consistent.

\subsubsection{Inference with Sparse Catalogues of Tracers}

As mentioned in section \ref{Cats}, we also performed inference with the structure formation and linear models on a catalogue of 500 tracers, in order to test the effectiveness of the method with sparser data. Corresponding results are presented in Appendix \ref{ap::scresults}. We find that the structure formation model is still able self-consistently to reconstruct the two-point statistics of the three fields, albeit with a greater variance than with 10000 tracers.
 
\subsection{Robustness to Model Mis-specification} \label{MM}
In this section, we deliberately impose model mis-specification by using a full $N$-body simulator and taking haloes from a $\textsc{rockstar}$ halo catalogue as the peculiar velocity tracers for our inference. Just as above, we assess our reconstructions qualitatively by comparing slices of the reconstructed fields, and quantitatively by looking at power and cross-power spectra.

\subsubsection{Sampler Warm-Up}
The warm-up behaviour of the HMC sampler is investigated in Appendix \ref{ap::hburnin}, which shows that the sampler starts to sample the desired region of the posterior after roughly 1000 samples, for either model. Furthermore, the autocorrelation length for the initial conditions within a typical voxel is roughly 200 samples.

\subsubsection{Reconstructed Fields} \label{sec::hfresults}

\begin{figure*}
\centering
\includegraphics[width=0.99\textwidth]{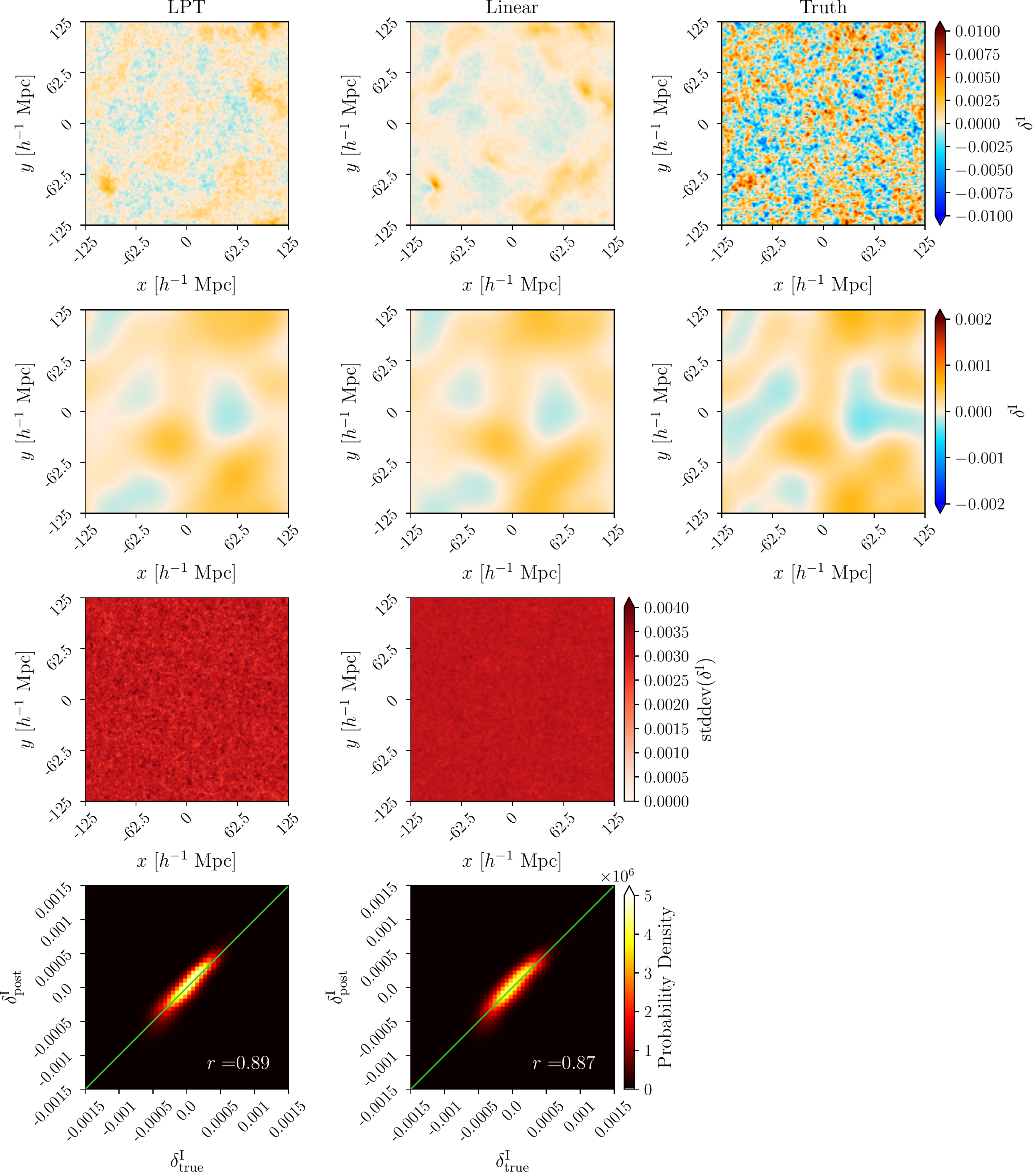}
\caption{\label{Hicslices} Top Row: from left to right, slices of the posterior mean initial conditions for the structure formation model (LPT) and linear model (Linear) using the halo mock catalogue, and finally the corresponding slice in the field from which the data are drawn (Truth), which we refer to as the `true' field. All are plotted with the same colour scaling. Second Row: the mean fields and true field plotted in the top row smoothed with a Gaussian filter at a $20\Mpch$ scale. Third Row: from left to right, slices of the posterior initial conditions standard deviation for the structure formation model and linear model, plotted with the same colour scaling.  Bottom Row: from left to right, normalised two-dimensional histograms of the initial conditions from samples against the true initial conditions for the structure formation model chain and the linear chain. We smooth each sample using the Gaussian filter described above and construct a histogram, and average the histograms along the chain. Visually, both models reproduce a smoothed version of the true initial conditions with a comparable variance. As both histograms are aligned with the green lines, the posterior distributions in each case are consistent with the distribution of the true initial conditions above the smoothing scale. The correlation coefficient is greater for the structure formation model, implying a greater concordance between this posterior and the true distribution.}
\end{figure*}

\begin{figure*}
\centering
\includegraphics[width=0.99\textwidth]{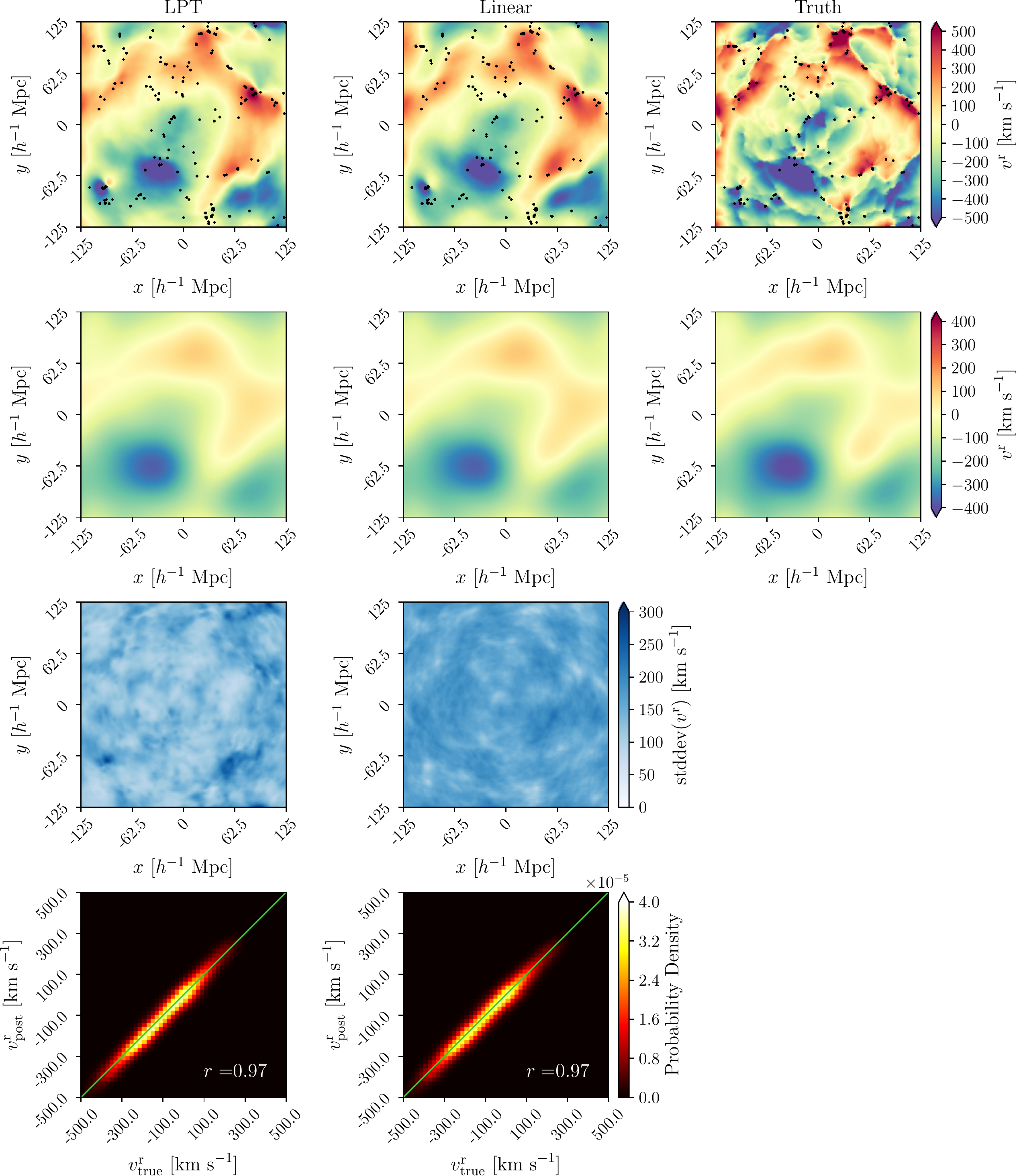}
\caption{\label{Hlosslices} Same as Fig. \ref{Hicslices} but for the posterior $z=0$ line-of-sight velocity field using the halo mock catalogue. The tracers with a perpendicular distance less than $1.9 \Mpch$ from this slice are plotted as black dots. Both models visually reconstruct the large scales true line-of-sight velocity field. The structure formation model does so with a generally lower variance. 
After smoothing the samples with a Gaussian filter with width $20\Mpch$, the histograms for both posterior distributions align with the green line, meaning that the posteriors are consistent with the true distribution for scales larger than this. 
}
\end{figure*}

\begin{figure*}
\centering
\includegraphics[width=0.99\textwidth]{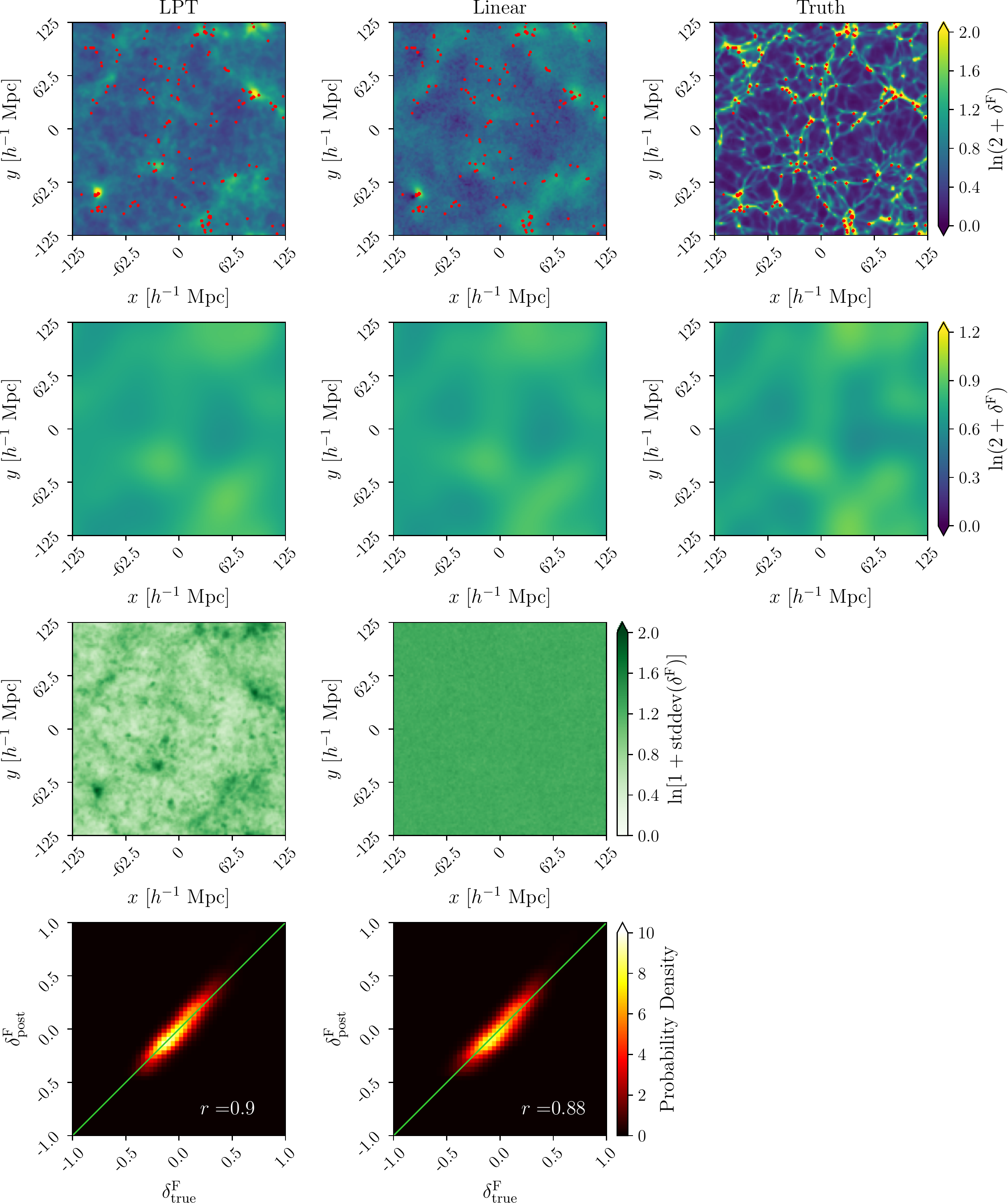}
\caption{\label{Hdenslices} Same as Fig. \ref{Hicslices} but for the posterior $z=0$ density contrast using the halo mock catalogue. The tracers with a perpendicular distance less than $1.9 \Mpch$ from this slice are plotted in red. The structure formation model reproduces a smoothed version of the true field, accurately capturing the large and intermediate scales. In contrast, the linear model does not accurately reproduce the intermediate and small scale power of the true field. Furthermore, the structure formation model reconstruction has a lower variance across the slice. The histograms for both models are aligned with the green line, showing that the distributions for the density contrast are consistent with the distribution of the true field at large scales. The correlation coefficient is greater for the structure formation model, implying a greater concordance between this posterior and the true distribution.}
\end{figure*}

As described in section \ref{Cats}, we model the non-linearity that is not captured by the data model as a random Gaussian scatter $\sigma_{\mathrm{MS}}$, which is added in quadrature to the value of $\sigma_\mathrm{noise}$ used in the likelihood. We tested values of $\sigma_{\mathrm{MS}}$ ranging from $0$ to $600$ $\mathrm{km} \mathrm{s}^{-1}$. Figures  \ref{Hicslices}, \ref{Hlosslices}, \ref{Hdenslices}, and \ref{HPspec} were all made using $\sigma_{\mathrm{MS}} = 300~ \mathrm{km} \mathrm{s}^{-1}$, such that $\sigma = \sqrt{150^{2} + 300^{2}}  \ \mathrm{km} \mathrm{s}^{-1}$. The behaviour discussed below was demonstrated for all $\sigma \geq 300 \mathrm{km} \mathrm{s}^{-1}$. 

We qualitatively investigate the reconstructed fields for the inference performed on the mock halo catalogue. In each case, on the top row, we show the central slice of the chain mean field for the structure formation model on the left, and the linear model in the middle. The true field used in the data generation process is on the right. In the second row, we show the fields from the top row smoothed using a Gaussian filter of $20\Mpch$ width, chosen as this is roughly the scale at which the cross-correlation between the chain mean initial conditions and the true initial conditions falls to 0.5 (see Figure \ref{HPspec}, top right panel). In the third row, we show the standard deviation of the chain in the same slice for the structure formation model on the left, and the linear model on the right. In the bottom row, we plot normalised two-dimensional histograms of the field from samples against the true field, for the structure formation model on the left, and the linear model on the right. We smooth each sample using the Gaussian filter described above and construct a histogram, and average the histograms along the chain. The correlation coefficient between the model posterior distribution and true distribution is given in each case.

In Figure \ref{Hicslices}, we show the reconstructed initial conditions. Here we see that, after smoothing at a $20\Mpch$ scale, both models capture some of the large-scale behaviour of the truth, but do not visually reproduce all of the small-scale features. Therefore, broadening the likelihood allows our structure formation model to capture the statistics of the true initial conditions in an unbiased way even in presence of model mis-specification, but reduces the predictive power of the method. The posteriors using the structure formation model and the linear model have broadly the same variance. As both histograms align with the green line, both posteriors are consistent with the posterior of the true initial conditions. Thus, we see that our structure formation model can handle model mis-specification in the data, at the level of the statistics of the initial conditions. 

In this test, we do not expect either the LPT or linear models to produce posterior distributions for the final density or final line-of-sight velocity field consistent with the truth at small scales. Indeed, samples of our chains are either LPT or linear theory realisations, whereas the true density and velocity fields are generated using a full $N$-body simulation, which captures the small-scale behaviour more realistically than both the LPT and linear models. In Figure \ref{Hlosslices}, we show the reconstructed $z=0$ line-of-sight velocity field. As can be observed, both models are able to reproduce the large scales of the true field. Because we must broaden the likelihood to account for the model mis-specification, neither model is able to capture the small-scale structure of the velocity field here. The structure formation reconstruction has a lower variance, however. The histograms for both models are aligned with the green line, showing that the distribution for both models are consistent with the distribution of the true field. 

In Figure \ref{Hdenslices} we show the reconstructed $z=0$ density contrast. The structure formation model reconstructs a smoothed version of the true density contrast, and accurately captures the intermediate and large-scale behaviour. The linear model captures the large-scale behaviour, but does not provide a good qualitative reconstruction of the truth due to excessive clustering at intermediate and small scales. The structure formation reconstructions have a lower variance across the slice. The histograms for both models are aligned with the green line, showing that the distributions for the density contrast are consistent with the true distribution at large scales. Similar to the initial conditions, the posterior for the structure formation model has a greater correlation coefficient than the posterior for the linear model, implying that the structure formation model posterior is in better correspondence with the truth.

\subsubsection{Power Spectra and Cross-Correlation Coefficient}

\begin{figure*}
\centering
\includegraphics[width=0.99\textwidth]{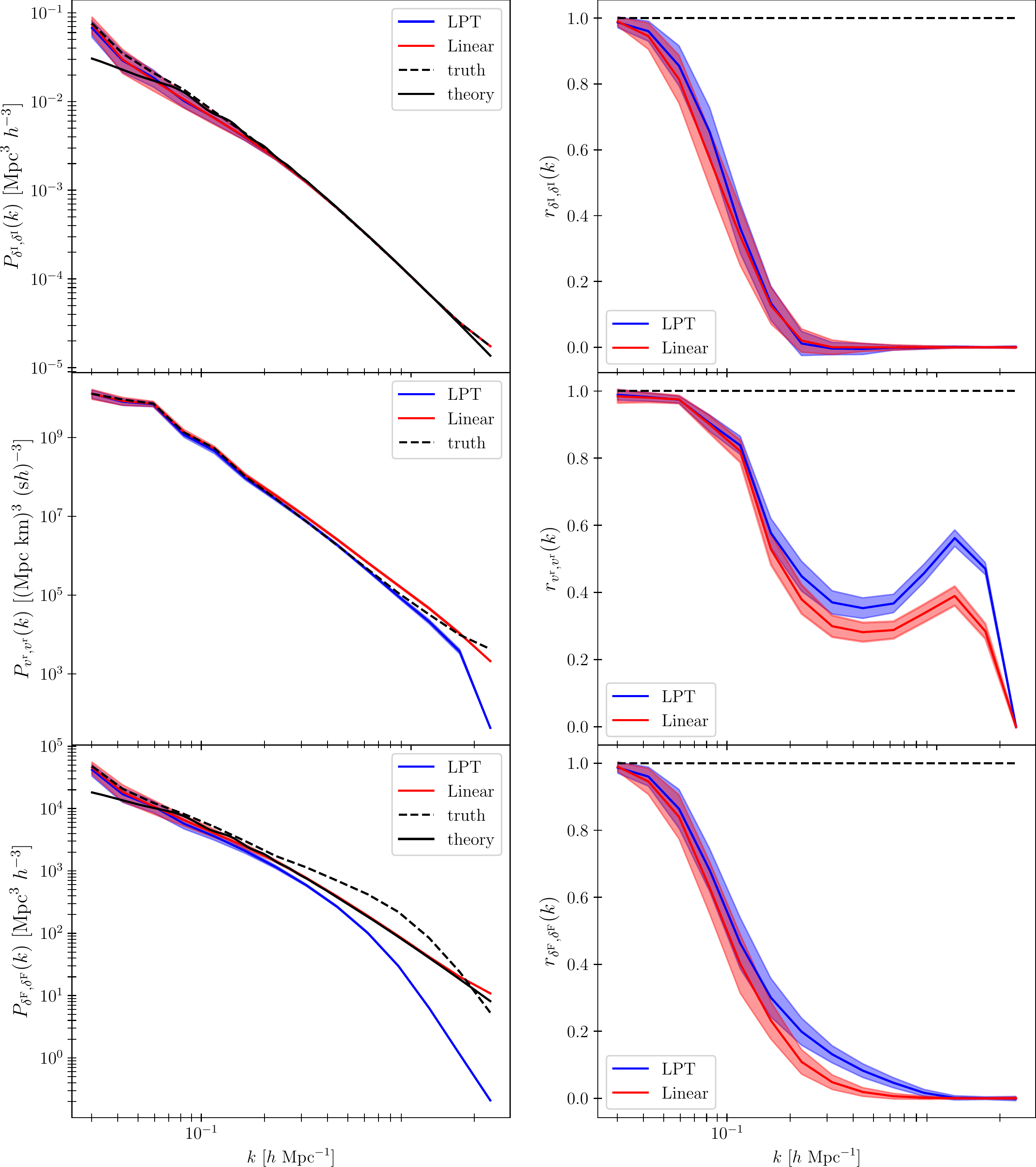}
\caption{\label{HPspec}
Same as Fig \ref{SCspec} but for the catalogue of 10000 mock halos. 
Both models are able to reproduce the power spectra of the true initial conditions at all scales.
Both models are also able to reproduce the large scales of the $z=0$ line-of-sight velocity field and $z=0$ density contrast. At small scales, the power spectra of both models are biased low (for $k \gtrsim 0.1h \ \mathrm{Mpc}^{-1}$), because neither the LPT nor the linear model capture the small-scale behaviour as realistically as the $N$-body simulation used to generate the true fields. The phase accuracy with the structure formation model is greater than or equal to that of the linear model for the initial conditions, $z=0$ density contrast, and $z=0$ line-of-sight velocity field at all scales.
}
\end{figure*}

In the left-hand column of Figure \ref{HPspec}, we plot the power spectra of samples produced from inference from the halo mock catalogue. For each field, we calculate the mean and $2 \sigma$ credible interval. We compare each case to the corresponding ‘true’ field used in the data generation process.

The top-left panel shows that both the structure formation and linear model reconstruct the initial conditions with power spectra consistent with the true initial conditions for all modes in our reconstruction box. 

We note that the linear model mean power spectrum is consistent with the truth here. As described in section \ref{sec::scresults}, the linear model samples have a generally larger variance than the structure formation model samples. This larger variance is the reason that the linear model appears consistent with the truth at the largest scale, as opposed to it having a greater accuracy than the structure formation model. As such, the structure formation model is able to reproduce the initial conditions at scales where we have sufficient tracers  if we broaden the likelihood. Therefore, we conclude that our method is robust to model mis-specification where the tracers are generated using a more realistic gravity model at the level of the statistics of the initial conditions.

The middle-left and bottom-left panels show that the reconstructed fields from both models are only able to reproduce the large scales of the true line-of-sight velocity and density contrast. The mean power spectrum is biased low for $k \gtrsim 0.1h \ \mathrm{Mpc}^{-1}$ for both models in the case of the $z=0$ density contrast and of the line-of-sight velocity field. 

Although broadening the likelihood allows our structure formation implementation to infer the initial conditions (target variables of the model) in an unbiased way, neither the LPT nor the linear model capture the small scale behaviour of the cosmic dark matter as accurately as the $N$-body model used to generate the ground truth, as discussed in section \ref{sec::hfresults}. Therefore, a different small-scale behaviour in chain samples and in the true $z=0$ density and line-of-sight velocity fields (latent variables of the model) is expected.

In the right-hand column, we plot the cross-correlation coefficient between the reconstructions and the truth. As in the left-hand column, we calculate the mean and $2 \sigma$ credible interval from each chain. For all three fields, the cross-correlation of the true field and the structure formation samples is greater than or equal to the cross-power spectrum of the true field and the linear model samples. Therefore, the reconstructed fields from the structure formation model have a phase accuracy greater than or equal to that of the linear model at all available scales.

We conclude that broadening the likelihood given in equation \eqref{VIRBIUS_likelihood} allows our structure formation model to be robust to model mis-specification arising from tracers being made using fully non-linear gravity and a halo finder. While our results demonstrate that using either linear or LPT models to evolve the cosmic dark matter allows us to reproduce the statistics of the initial conditions, we note that we expect that using a more physical data model will allow us to use a lower value of $\sigma_\mathrm{NL}$ and obtain better reconstructions of the final fields.

\section{Discussion and Conclusions} \label{sec::Conc}

\subsection{Summary}
We have adapted the Bayesian reconstruction algorithm $\textsc{borg}$ to perform field-based physical inference from peculiar velocity tracers. \textcolor{black}{The aim of this work was to improve the physical modelling aspect of the inference using peculiar velocity tracers.} The method employs a Gaussian prior for the initial conditions, and makes use of a structure formation model to link these initial conditions to late-time observations. It uses a Gaussian likelihood for the line-of-sight component of the velocity field, evaluated at the positions of tracers. In this paper, we assume that the data are composed of peculiar velocity tracers, i.e. galaxies with distance indicators for which the error in the position is negligible. As a result, we are able to infer the initial density field from peculiar velocity tracers. As derived products, we also obtain physical reconstructions of the $z=0$ density field and velocity field. 

We compared our approach to the standard reconstruction framework, used in particular in Wiener filtering, which assumes a Gaussian-distributed velocity field linearly linked to the initial conditions. We found qualitative and quantitative (at the level of one- and two-point statistics) improvements in the reconstruction of density and velocity fields. In a self-consistency test of our method on a set of mock tracers drawn from the data model, we showed that physical reconstructions visually reproduce the true fields (see Figs. \ref{SCicslices}, \ref{SClosslices}, \ref{SCdenslices}, and \ref{SCangslices}) and are accurate at the level of two-point statistics (Fig. \ref{SCspec}). In contrast, linear modelling fails at small scales. 

This test demonstrates that the method is self-consistent and solves a potential shortcoming of the previous state-of-the-art technique for velocity field reconstruction, \textcolor{black}{namely that the statistics of the reconstructed fields are biased when using a linear model.}

Model mis-specification is a major challenge that field-based cosmological inference approaches have to face in order to obtain unbiased results in the presence of  incomplete physical modelling and/or unknown systematics \citep[see][]{Porqueres2019,2019arXiv190906396L,LavauxJasche2021}. In this paper, we modelled the theoretical uncertainty arising from model mis-specification as a Gaussian random scatter added in quadrature to the observational error, increasing the width of the Gaussian likelihood. We tested our method on a more realistic set of mock tracers, drawn from a $\textsc{rockstar}$ halo catalogue made using a $\textsc{gadget2}$ snapshot. In this scenario, we found that our approach is able to provide an unbiased inference of the target fields, albeit with increased variance.

\subsection{Future Improvements} \label{sec::futurework}
There are several improvements that we aim to implement in future work. Independently of the physical model, observational effects such as survey geometries and radial selection functions can be included in the Bayesian hierarchical model \citep[see][]{2016MNRAS.457..172L,2019MNRAS.488.5438G,Boruah2021}.

Furthermore, throughout the paper, we have assumed that Malmquist bias can be neglected, and thus that we exactly know the distance to each tracer. This assumption allows a simplified modelling of peculiar velocity tracers, valid for low redshifts only. 
\textcolor{black}{In future work, to prepare our method for real data applications, we will relax this assumption by sampling the luminosity distance of tracers, conditional on their distance moduli and redshift, similar to \citet{2016MNRAS.457..172L,2019MNRAS.488.5438G,2022MNRAS.513.5148V} and \citet{Boruah2021}. }

We will also explore the use of a more accurate non-linear structure formation model. 
$\textsc{borg}$ has previously been run using second-order LPT \citep{2015JCAP...01..036J,2016MNRAS.455.3169L,2019arXiv190906396L}, as well as a full particle-mesh code \citep{2019A&A...625A..64J}. Each of these are more physical descriptions of gravitational dynamics than the Zel'dovich approximation used in this work, and so should allow us to probe further into the non-linear regime. We expect that using a better gravity model will further increase the accuracy of reconstructions at small scales. \textcolor{black}{We also expect that using a better gravity model will allow us to perform reconstructions with a smaller value of $\sigma_{\mathrm{MS}}$.}

Furthermore, we will also explore different models for the velocity field. In this work, we have used the linearised continuity equation (equation \eqref{EPTvel}). We will go beyond this description in the future, using models that better encapsulate the non-linear evolution of dark matter in phase space. These include kernel methods based on the velocity of simulation particles, as well as the simplex-in-cell estimator \citep{HahnAnguloAbel2015,2017JCAP...06..049L}.

\subsection{Conclusions}
To summarise, we have introduced and validated a physical field-based inference method for reconstructing cosmological fields constrained by radial peculiar velocity observations. This method uses a structure formation model to link the initial conditions to late-time density and velocity fields, allowing us to reconstruct all of these from observations. Introducing structure formation in the model accounts for the non-linear gravitational evolution of cosmic matter, and allows for more accurate velocity field reconstructions than can be achieved with a linear model.

\section*{Acknowledgements}
We thank Eleni Tsaprazi, Deaglan Bartlett, and Natalia Porqueres for their helpful comments. This work was supported by the Simons Collaboration on ``Learning the Universe''. This work was done within the Aquila Consortium\footnote{\url{https://aquila-consortium.org}}.
FL acknowledges funding from the Imperial College London Research Fellowship Scheme and the STFC Ernest Rutherford Fellowship, Grant ST/V004239/1. GL acknowledges support by the ANR BIG4 project, grant ANR-16-CE23-0002 of the French Agence Nationale de la Recherche. JJ acknowledges support by the Swedish Research Council (VR) under the project 2020-05143 -- ``Deciphering the Dynamics of Cosmic Structure".

\section*{Data Availability}
The data underlying this article will be shared on reasonable request to the corresponding author.


\bibliographystyle{mnras}
\bibliography{biblio}

\begin{thebibliography}{}
\makeatletter
\relax
\def\mn@urlcharsother{\let\do\@makeother \do\$\do\&\do\#\do\^\do\_\do\%\do\~}
\def\mn@doi{\begingroup\mn@urlcharsother \@ifnextchar [ {\mn@doi@}
  {\mn@doi@[]}}
\def\mn@doi@[#1]#2{\def\@tempa{#1}\ifx\@tempa\@empty \href
  {http://dx.doi.org/#2} {doi:#2}\else \href {http://dx.doi.org/#2} {#1}\fi
  \endgroup}
\def\mn@eprint#1#2{\mn@eprint@#1:#2::\@nil}
\def\mn@eprint@arXiv#1{\href {http://arxiv.org/abs/#1} {{\tt arXiv:#1}}}
\def\mn@eprint@dblp#1{\href {http://dblp.uni-trier.de/rec/bibtex/#1.xml}
  {dblp:#1}}
\def\mn@eprint@#1:#2:#3:#4\@nil{\def\@tempa {#1}\def\@tempb {#2}\def\@tempc
  {#3}\ifx \@tempc \@empty \let \@tempc \@tempb \let \@tempb \@tempa \fi \ifx
  \@tempb \@empty \def\@tempb {arXiv}\fi \@ifundefined
  {mn@eprint@\@tempb}{\@tempb:\@tempc}{\expandafter \expandafter \csname
  mn@eprint@\@tempb\endcsname \expandafter{\@tempc}}}

\bibitem[\protect\citeauthoryear{{Ata}, {Kitaura}, {Lee}, {Lemaux}, {Kashino},
  {Cucciati}, {Hern{\'a}ndez-S{\'a}nchez}  \& {Le F{\`e}vre}}{{Ata}
  et~al.}{2021}]{2021MNRAS.500.3194A}
{Ata} M.,  {Kitaura} F.-S.,  {Lee} K.-G.,  {Lemaux} B.~C.,  {Kashino} D.,
  {Cucciati} O.,  {Hern{\'a}ndez-S{\'a}nchez} M.,   {Le F{\`e}vre} O.,  2021,
  \mn@doi [\mnras] {10.1093/mnras/staa3318}, \href
  {https://ui.adsabs.harvard.edu/abs/2021MNRAS.500.3194A} {500, 3194}

\bibitem[\protect\citeauthoryear{Bartlett, Desmond  \& Ferreira}{Bartlett
  et~al.}{2021}]{PhysRevD.103.023523}
Bartlett D.~J.,  Desmond H.,   Ferreira P.~G.,  2021, \mn@doi [Phys. Rev. D]
  {10.1103/PhysRevD.103.023523}, 103, 023523

\bibitem[\protect\citeauthoryear{{Behroozi}, {Wechsler}  \& {Wu}}{{Behroozi}
  et~al.}{2013}]{2013ApJ...762..109B}
{Behroozi} P.~S.,  {Wechsler} R.~H.,   {Wu} H.-Y.,  2013, \mn@doi [\apj]
  {10.1088/0004-637X/762/2/109}, \href
  {https://ui.adsabs.harvard.edu/abs/2013ApJ...762..109B} {762, 109}

\bibitem[\protect\citeauthoryear{{Bernardeau}, {Colombi}, {Gazta{\~n}aga}  \&
  {Scoccimarro}}{{Bernardeau} et~al.}{2002}]{2002PhR...367....1B}
{Bernardeau} F.,  {Colombi} S.,  {Gazta{\~n}aga} E.,   {Scoccimarro} R.,  2002,
  \mn@doi [\physrep] {10.1016/S0370-1573(02)00135-7}, \href
  {https://ui.adsabs.harvard.edu/abs/2002PhR...367....1B} {367, 1}

\bibitem[\protect\citeauthoryear{{Bertschinger}}{{Bertschinger}}{2011}]{2011RSPTA.369.4947B}
{Bertschinger} E.,  2011, \mn@doi [Philosophical Transactions of the Royal
  Society of London Series A] {10.1098/rsta.2011.0369}, \href
  {https://ui.adsabs.harvard.edu/abs/2011RSPTA.369.4947B} {369, 4947}

\bibitem[\protect\citeauthoryear{{Bertschinger} \& {Dekel}}{{Bertschinger} \&
  {Dekel}}{1989}]{1989ApJ...336L...5B}
{Bertschinger} E.,  {Dekel} A.,  1989, \mn@doi [\apjl] {10.1086/185348}, \href
  {https://ui.adsabs.harvard.edu/abs/1989ApJ...336L...5B} {336, L5}

\bibitem[\protect\citeauthoryear{{Boruah}, {Hudson}  \& {Lavaux}}{{Boruah}
  et~al.}{2020}]{2020MNRAS.498.2703B}
{Boruah} S.~S.,  {Hudson} M.~J.,   {Lavaux} G.,  2020, \mn@doi [\mnras]
  {10.1093/mnras/staa2485}, \href
  {https://ui.adsabs.harvard.edu/abs/2020MNRAS.498.2703B} {498, 2703}

\bibitem[\protect\citeauthoryear{Boruah, Lavaux  \& Hudson}{Boruah
  et~al.}{2021}]{Boruah2021}
Boruah S.~S.,  Lavaux G.,   Hudson M.~J.,  2021, arXiv e-prints, p.
  arXiv:2111.15535

\bibitem[\protect\citeauthoryear{{Buehlmann} \& {Hahn}}{{Buehlmann} \&
  {Hahn}}{2019}]{2019MNRAS.487..228B}
{Buehlmann} M.,  {Hahn} O.,  2019, \mn@doi [\mnras] {10.1093/mnras/stz1243},
  \href {https://ui.adsabs.harvard.edu/abs/2019MNRAS.487..228B} {487, 228}

\bibitem[\protect\citeauthoryear{{Campbell} et~al.,}{{Campbell}
  et~al.}{2014}]{2014MNRAS.443.1231C}
{Campbell} L.~A.,  et~al., 2014, \mn@doi [\mnras] {10.1093/mnras/stu1198},
  \href {https://ui.adsabs.harvard.edu/abs/2014MNRAS.443.1231C} {443, 1231}

\bibitem[\protect\citeauthoryear{{Carrick}, {Turnbull}, {Lavaux}  \&
  {Hudson}}{{Carrick} et~al.}{2015}]{2015MNRAS.450..317C}
{Carrick} J.,  {Turnbull} S.~J.,  {Lavaux} G.,   {Hudson} M.~J.,  2015, \mn@doi
  [\mnras] {10.1093/mnras/stv547}, \href
  {https://ui.adsabs.harvard.edu/abs/2015MNRAS.450..317C} {450, 317}

\bibitem[\protect\citeauthoryear{{Courtois}, {Hoffman}, {Tully}  \&
  {Gottl{\"o}ber}}{{Courtois} et~al.}{2012}]{2012ApJ...744...43C}
{Courtois} H.~M.,  {Hoffman} Y.,  {Tully} R.~B.,   {Gottl{\"o}ber} S.,  2012,
  \mn@doi [\apj] {10.1088/0004-637X/744/1/43}, \href
  {https://ui.adsabs.harvard.edu/abs/2012ApJ...744...43C} {744, 43}

\bibitem[\protect\citeauthoryear{{Dekel}, {Bertschinger}  \& {Faber}}{{Dekel}
  et~al.}{1990}]{1990ApJ...364..349D}
{Dekel} A.,  {Bertschinger} E.,   {Faber} S.~M.,  1990, \mn@doi [\apj]
  {10.1086/169418}, \href
  {https://ui.adsabs.harvard.edu/abs/1990ApJ...364..349D} {364, 349}

\bibitem[\protect\citeauthoryear{{Dekel}, {Eldar}, {Kolatt}, {Yahil},
  {Willick}, {Faber}, {Courteau}  \& {Burstein}}{{Dekel}
  et~al.}{1999}]{1999ApJ...522....1D}
{Dekel} A.,  {Eldar} A.,  {Kolatt} T.,  {Yahil} A.,  {Willick} J.~A.,  {Faber}
  S.~M.,  {Courteau} S.,   {Burstein} D.,  1999, \mn@doi [\apj]
  {10.1086/307636}, \href
  {https://ui.adsabs.harvard.edu/abs/1999ApJ...522....1D} {522, 1}

\bibitem[\protect\citeauthoryear{Desmond, Ferreira, Lavaux  \& Jasche}{Desmond
  et~al.}{2018}]{PhysRevD.98.083010}
Desmond H.,  Ferreira P.~G.,  Lavaux G.,   Jasche J.,  2018, \mn@doi [Phys.
  Rev. D] {10.1103/PhysRevD.98.083010}, 98, 083010

\bibitem[\protect\citeauthoryear{{Doumler}, {Hoffman}, {Courtois}  \&
  {Gottl{\"o}ber}}{{Doumler} et~al.}{2013a}]{2013MNRAS.430..888D}
{Doumler} T.,  {Hoffman} Y.,  {Courtois} H.,   {Gottl{\"o}ber} S.,  2013a,
  \mn@doi [\mnras] {10.1093/mnras/sts613}, \href
  {https://ui.adsabs.harvard.edu/abs/2013MNRAS.430..888D} {430, 888}

\bibitem[\protect\citeauthoryear{{Doumler}, {Courtois}, {Gottl{\"o}ber}  \&
  {Hoffman}}{{Doumler} et~al.}{2013b}]{2013MNRAS.430..902D}
{Doumler} T.,  {Courtois} H.,  {Gottl{\"o}ber} S.,   {Hoffman} Y.,  2013b,
  \mn@doi [\mnras] {10.1093/mnras/sts603}, \href
  {https://ui.adsabs.harvard.edu/abs/2013MNRAS.430..902D} {430, 902}

\bibitem[\protect\citeauthoryear{{Doumler}, {Gottl{\"o}ber}, {Hoffman}  \&
  {Courtois}}{{Doumler} et~al.}{2013c}]{2013MNRAS.430..912D}
{Doumler} T.,  {Gottl{\"o}ber} S.,  {Hoffman} Y.,   {Courtois} H.,  2013c,
  \mn@doi [\mnras] {10.1093/mnras/sts614}, \href
  {https://ui.adsabs.harvard.edu/abs/2013MNRAS.430..912D} {430, 912}

\bibitem[\protect\citeauthoryear{{Duane}, {Kennedy}, {Pendleton}  \&
  {Roweth}}{{Duane} et~al.}{1987}]{1987PhLB..195..216D}
{Duane} S.,  {Kennedy} A.~D.,  {Pendleton} B.~J.,   {Roweth} D.,  1987, \mn@doi
  [Physics Letters B] {10.1016/0370-2693(87)91197-X}, \href
  {https://ui.adsabs.harvard.edu/abs/1987PhLB..195..216D} {195, 216}

\bibitem[\protect\citeauthoryear{{Eisenstein} \& {Hu}}{{Eisenstein} \&
  {Hu}}{1998}]{1998ApJ...496..605E}
{Eisenstein} D.~J.,  {Hu} W.,  1998, \mn@doi [\apj] {10.1086/305424}, \href
  {https://ui.adsabs.harvard.edu/abs/1998ApJ...496..605E} {496, 605}

\bibitem[\protect\citeauthoryear{{Eisenstein} \& {Hu}}{{Eisenstein} \&
  {Hu}}{1999}]{1999ApJ...511....5E}
{Eisenstein} D.~J.,  {Hu} W.,  1999, \mn@doi [\apj] {10.1086/306640}, \href
  {https://ui.adsabs.harvard.edu/abs/1999ApJ...511....5E} {511, 5}

\bibitem[\protect\citeauthoryear{{Graziani}, {Courtois}, {Lavaux}, {Hoffman},
  {Tully}, {Copin}  \& {Pomar{\`e}de}}{{Graziani}
  et~al.}{2019}]{2019MNRAS.488.5438G}
{Graziani} R.,  {Courtois} H.~M.,  {Lavaux} G.,  {Hoffman} Y.,  {Tully} R.~B.,
  {Copin} Y.,   {Pomar{\`e}de} D.,  2019, \mn@doi [\mnras]
  {10.1093/mnras/stz078}, \href
  {https://ui.adsabs.harvard.edu/abs/2019MNRAS.488.5438G} {488, 5438}

\bibitem[\protect\citeauthoryear{{Hahn}, {Angulo}  \& {Abel}}{{Hahn}
  et~al.}{2015}]{HahnAnguloAbel2015}
{Hahn} O.,  {Angulo} R.~E.,   {Abel} T.,  2015, \mn@doi [\mnras]
  {10.1093/mnras/stv2179}, \href
  {http://adsabs.harvard.edu/abs/2015MNRAS.454.3920H} {454, 3920}

\bibitem[\protect\citeauthoryear{{He{\ss}}, {Kitaura}  \&
  {Gottl{\"o}ber}}{{He{\ss}} et~al.}{2013}]{2013MNRAS.435.2065H}
{He{\ss}} S.,  {Kitaura} F.-S.,   {Gottl{\"o}ber} S.,  2013, \mn@doi [\mnras]
  {10.1093/mnras/stt1428}, \href
  {https://ui.adsabs.harvard.edu/abs/2013MNRAS.435.2065H} {435, 2065}

\bibitem[\protect\citeauthoryear{Hockney \& Eastwood}{Hockney \&
  Eastwood}{1981}]{CiC}
Hockney R.~W.,  Eastwood J.~W.,  1981, Computer Simulation Using Particles..
McGraw-Hill

\bibitem[\protect\citeauthoryear{{Hoffman} \& {Ribak}}{{Hoffman} \&
  {Ribak}}{1991}]{Hoffman1991}
{Hoffman} Y.,  {Ribak} E.,  1991, \mn@doi [\apjl] {10.1086/186160}, \href
  {http://adsabs.harvard.edu/abs/1991ApJ...380L...5H} {380, L5}

\bibitem[\protect\citeauthoryear{{Hoffman}, {Courtois}  \& {Tully}}{{Hoffman}
  et~al.}{2015}]{2015MNRAS.449.4494H}
{Hoffman} Y.,  {Courtois} H.~M.,   {Tully} R.~B.,  2015, \mn@doi [\mnras]
  {10.1093/mnras/stv615}, \href
  {https://ui.adsabs.harvard.edu/abs/2015MNRAS.449.4494H} {449, 4494}

\bibitem[\protect\citeauthoryear{{Hoffman} et~al.,}{{Hoffman}
  et~al.}{2018}]{2018NatAs...2..680H}
{Hoffman} Y.,  et~al., 2018, \mn@doi [Nature Astronomy]
  {10.1038/s41550-018-0502-4}, \href
  {https://ui.adsabs.harvard.edu/abs/2018NatAs...2..680H} {2, 680}

\bibitem[\protect\citeauthoryear{{Hoffman}, {Nusser}, {Valade}, {Libeskind}  \&
  {Tully}}{{Hoffman} et~al.}{2021}]{Hoffman2021}
{Hoffman} Y.,  {Nusser} A.,  {Valade} A.,  {Libeskind} N.~I.,   {Tully} R.~B.,
  2021, \mn@doi [\mnras] {10.1093/mnras/stab1457}, \href
  {https://ui.adsabs.harvard.edu/abs/2021MNRAS.505.3380H} {505, 3380}

\bibitem[\protect\citeauthoryear{{Hong}, {Jeong}, {Hwang}  \& {Kim}}{{Hong}
  et~al.}{2021}]{2021ApJ...913...76H}
{Hong} S.~E.,  {Jeong} D.,  {Hwang} H.~S.,   {Kim} J.,  2021, \mn@doi [\apj]
  {10.3847/1538-4357/abf040}, \href
  {https://ui.adsabs.harvard.edu/abs/2021ApJ...913...76H} {913, 76}

\bibitem[\protect\citeauthoryear{{Howlett} et~al.,}{{Howlett}
  et~al.}{2017}]{2017MNRAS.471.3135H}
{Howlett} C.,  et~al., 2017, \mn@doi [\mnras] {10.1093/mnras/stx1521}, \href
  {https://ui.adsabs.harvard.edu/abs/2017MNRAS.471.3135H} {471, 3135}

\bibitem[\protect\citeauthoryear{{Jasche} \& {Kitaura}}{{Jasche} \&
  {Kitaura}}{2010}]{2010MNRAS.407...29J}
{Jasche} J.,  {Kitaura} F.~S.,  2010, \mn@doi [\mnras]
  {10.1111/j.1365-2966.2010.16897.x}, \href
  {https://ui.adsabs.harvard.edu/abs/2010MNRAS.407...29J} {407, 29}

\bibitem[\protect\citeauthoryear{{Jasche} \& {Lavaux}}{{Jasche} \&
  {Lavaux}}{2019}]{2019A&A...625A..64J}
{Jasche} J.,  {Lavaux} G.,  2019, \mn@doi [\aap] {10.1051/0004-6361/201833710},
  \href {https://ui.adsabs.harvard.edu/abs/2019A&A...625A..64J} {625, A64}

\bibitem[\protect\citeauthoryear{{Jasche} \& {Wandelt}}{{Jasche} \&
  {Wandelt}}{2013}]{2013MNRAS.432..894J}
{Jasche} J.,  {Wandelt} B.~D.,  2013, \mn@doi [\mnras] {10.1093/mnras/stt449},
  \href {https://ui.adsabs.harvard.edu/abs/2013MNRAS.432..894J} {432, 894}

\bibitem[\protect\citeauthoryear{{Jasche}, {Leclercq}  \& {Wandelt}}{{Jasche}
  et~al.}{2015}]{2015JCAP...01..036J}
{Jasche} J.,  {Leclercq} F.,   {Wandelt} B.~D.,  2015, \mn@doi [\jcap]
  {10.1088/1475-7516/2015/01/036}, \href
  {https://ui.adsabs.harvard.edu/abs/2015JCAP...01..036J} {2015, 036}

\bibitem[\protect\citeauthoryear{Jaynes}{Jaynes}{2003}]{jaynes03}
Jaynes E.~T.,  2003, Probability theory: The logic of science.
Cambridge University Press, Cambridge

\bibitem[\protect\citeauthoryear{Kitaura \& Enßlin}{Kitaura \&
  Enßlin}{2008}]{10.1111/j.1365-2966.2008.13341.x}
Kitaura F.~S.,  Enßlin T.~A.,  2008, \mn@doi [Monthly Notices of the Royal
  Astronomical Society] {10.1111/j.1365-2966.2008.13341.x}, 389, 497

\bibitem[\protect\citeauthoryear{{Kitaura}, {Ata}, {Rodr{\'\i}guez-Torres},
  {Hern{\'a}ndez-S{\'a}nchez}, {Balaguera-Antol{\'\i}nez}  \&
  {Yepes}}{{Kitaura} et~al.}{2021}]{2021MNRAS.502.3456K}
{Kitaura} F.-S.,  {Ata} M.,  {Rodr{\'\i}guez-Torres} S.~A.,
  {Hern{\'a}ndez-S{\'a}nchez} M.,  {Balaguera-Antol{\'\i}nez} A.,   {Yepes} G.,
   2021, \mn@doi [\mnras] {10.1093/mnras/staa3774}, \href
  {https://ui.adsabs.harvard.edu/abs/2021MNRAS.502.3456K} {502, 3456}

\bibitem[\protect\citeauthoryear{{Kolatt}, {Dekel}, {Ganon}  \&
  {Willick}}{{Kolatt} et~al.}{1996}]{1996ApJ...458..419K}
{Kolatt} T.,  {Dekel} A.,  {Ganon} G.,   {Willick} J.~A.,  1996, \mn@doi [\apj]
  {10.1086/176826}, \href
  {https://ui.adsabs.harvard.edu/abs/1996ApJ...458..419K} {458, 419}

\bibitem[\protect\citeauthoryear{{Kourkchi} et~al.,}{{Kourkchi}
  et~al.}{2020}]{2020ApJ...902..145K}
{Kourkchi} E.,  et~al., 2020, \mn@doi [\apj] {10.3847/1538-4357/abb66b}, \href
  {https://ui.adsabs.harvard.edu/abs/2020ApJ...902..145K} {902, 145}

\bibitem[\protect\citeauthoryear{{Lavaux}}{{Lavaux}}{2016}]{2016MNRAS.457..172L}
{Lavaux} G.,  2016, \mn@doi [\mnras] {10.1093/mnras/stv2915}, \href
  {https://ui.adsabs.harvard.edu/abs/2016MNRAS.457..172L} {457, 172}

\bibitem[\protect\citeauthoryear{{Lavaux} \& {Jasche}}{{Lavaux} \&
  {Jasche}}{2016}]{2016MNRAS.455.3169L}
{Lavaux} G.,  {Jasche} J.,  2016, \mn@doi [\mnras] {10.1093/mnras/stv2499},
  \href {https://ui.adsabs.harvard.edu/abs/2016MNRAS.455.3169L} {455, 3169}

\bibitem[\protect\citeauthoryear{{Lavaux} \& {Jasche}}{{Lavaux} \&
  {Jasche}}{2021}]{LavauxJasche2021}
{Lavaux} G.,  {Jasche} J.,  2021, arXiv e-prints, \href
  {https://ui.adsabs.harvard.edu/abs/2021arXiv210412992L} {p. arXiv:2104.12992}

\bibitem[\protect\citeauthoryear{{Lavaux}, {Jasche}  \& {Leclercq}}{{Lavaux}
  et~al.}{2019}]{2019arXiv190906396L}
{Lavaux} G.,  {Jasche} J.,   {Leclercq} F.,  2019, arXiv e-prints, \href
  {https://ui.adsabs.harvard.edu/abs/2019arXiv190906396L} {p. arXiv:1909.06396}

\bibitem[\protect\citeauthoryear{{Leclercq}, {Jasche}  \& {Wandelt}}{{Leclercq}
  et~al.}{2015}]{2015JCAP...06..015L}
{Leclercq} F.,  {Jasche} J.,   {Wandelt} B.,  2015, \mn@doi [\jcap]
  {10.1088/1475-7516/2015/06/015}, \href
  {https://ui.adsabs.harvard.edu/abs/2015JCAP...06..015L} {2015, 015}

\bibitem[\protect\citeauthoryear{{Leclercq}, {Jasche}, {Lavaux}, {Wandelt}  \&
  {Percival}}{{Leclercq} et~al.}{2017}]{2017JCAP...06..049L}
{Leclercq} F.,  {Jasche} J.,  {Lavaux} G.,  {Wandelt} B.,   {Percival} W.,
  2017, \mn@doi [\jcap] {10.1088/1475-7516/2017/06/049}, \href
  {https://ui.adsabs.harvard.edu/abs/2017JCAP...06..049L} {2017, 049}

\bibitem[\protect\citeauthoryear{Libeskind et~al.,}{Libeskind
  et~al.}{2020}]{10.1093/mnras/staa2541}
Libeskind N.~I.,  et~al., 2020, \mn@doi [Monthly Notices of the Royal
  Astronomical Society] {10.1093/mnras/staa2541}, 498, 2968

\bibitem[\protect\citeauthoryear{{Lynden-Bell}, {Faber}, {Burstein}, {Davies},
  {Dressler}, {Terlevich}  \& {Wegner}}{{Lynden-Bell}
  et~al.}{1988}]{1988ApJ...326...19L}
{Lynden-Bell} D.,  {Faber} S.~M.,  {Burstein} D.,  {Davies} R.~L.,  {Dressler}
  A.,  {Terlevich} R.~J.,   {Wegner} G.,  1988, \mn@doi [\apj]
  {10.1086/166066}, \href
  {https://ui.adsabs.harvard.edu/abs/1988ApJ...326...19L} {326, 19}

\bibitem[\protect\citeauthoryear{Neal}{Neal}{1993}]{Neal1993}
Neal R.~M.,  1993, Probabilistic inference using Markov chain Monte Carlo
  methods.
Tech. Rep. CRG-T

\bibitem[\protect\citeauthoryear{Neal}{Neal}{1996}]{Neal1996}
Neal R.~M.,  1996, Bayesian Learning for Neural Networks (Lecture Notes in
  Statistics).
1st edn. (Springer)

\bibitem[\protect\citeauthoryear{{Nguyen}, {Schmidt}, {Lavaux}  \&
  {Jasche}}{{Nguyen} et~al.}{2021}]{2021JCAP...03..058N}
{Nguyen} N.-M.,  {Schmidt} F.,  {Lavaux} G.,   {Jasche} J.,  2021, \mn@doi
  [\jcap] {10.1088/1475-7516/2021/03/058}, \href
  {https://ui.adsabs.harvard.edu/abs/2021JCAP...03..058N} {2021, 058}

\bibitem[\protect\citeauthoryear{{Pizzuti}, {Saltas}, {Casas}, {Amendola}  \&
  {Biviano}}{{Pizzuti} et~al.}{2019}]{2019MNRAS.486..596P}
{Pizzuti} L.,  {Saltas} I.~D.,  {Casas} S.,  {Amendola} L.,   {Biviano} A.,
  2019, \mn@doi [\mnras] {10.1093/mnras/stz825}, \href
  {https://ui.adsabs.harvard.edu/abs/2019MNRAS.486..596P} {486, 596}

\bibitem[\protect\citeauthoryear{{Planck Collaboration} et~al.,}{{Planck
  Collaboration} et~al.}{2020a}]{2020A&A...641A...6P}
{Planck Collaboration} et~al., 2020a, \mn@doi [\aap]
  {10.1051/0004-6361/201833910}, \href
  {https://ui.adsabs.harvard.edu/abs/2020A&A...641A...6P} {641, A6}

\bibitem[\protect\citeauthoryear{{Planck Collaboration} et~al.,}{{Planck
  Collaboration} et~al.}{2020b}]{2020A&A...641A...9P}
{Planck Collaboration} et~al., 2020b, \mn@doi [\aap]
  {10.1051/0004-6361/201935891}, \href
  {https://ui.adsabs.harvard.edu/abs/2020A&A...641A...9P} {641, A9}

\bibitem[\protect\citeauthoryear{Porqueres, Kodi~Ramanah, Jasche  \&
  Lavaux}{Porqueres et~al.}{2019}]{Porqueres2019}
Porqueres N.,  Kodi~Ramanah D.,  Jasche J.,   Lavaux G.,  2019, \mn@doi [\aap]
  {10.1051/0004-6361/201834844}, 624, A115

\bibitem[\protect\citeauthoryear{{Porqueres}, {Heavens}, {Mortlock}  \&
  {Lavaux}}{{Porqueres} et~al.}{2021}]{2021MNRAS.502.3035P}
{Porqueres} N.,  {Heavens} A.,  {Mortlock} D.,   {Lavaux} G.,  2021, \mn@doi
  [\mnras] {10.1093/mnras/stab204}, \href
  {https://ui.adsabs.harvard.edu/abs/2021MNRAS.502.3035P} {502, 3035}

\bibitem[\protect\citeauthoryear{{Porqueres}, {Heavens}, {Mortlock}  \&
  {Lavaux}}{{Porqueres} et~al.}{2022}]{2022MNRAS.509.3194P}
{Porqueres} N.,  {Heavens} A.,  {Mortlock} D.,   {Lavaux} G.,  2022, \mn@doi
  [\mnras] {10.1093/mnras/stab3234}, \href
  {https://ui.adsabs.harvard.edu/abs/2022MNRAS.509.3194P} {509, 3194}

\bibitem[\protect\citeauthoryear{{Sereno}}{{Sereno}}{2015}]{2015MNRAS.450.3665S}
{Sereno} M.,  2015, \mn@doi [\mnras] {10.1093/mnras/stu2505}, \href
  {https://ui.adsabs.harvard.edu/abs/2015MNRAS.450.3665S} {450, 3665}

\bibitem[\protect\citeauthoryear{{Sorce}}{{Sorce}}{2015}]{Sorce2015}
{Sorce} J.~G.,  2015, \mn@doi [\mnras] {10.1093/mnras/stv760}, \href
  {https://ui.adsabs.harvard.edu/abs/2015MNRAS.450.2644S} {450, 2644}

\bibitem[\protect\citeauthoryear{Sorce \& Tempel}{Sorce \&
  Tempel}{2018}]{10.1093/mnras/sty505}
Sorce J.~G.,  Tempel E.,  2018, \mn@doi [Monthly Notices of the Royal
  Astronomical Society] {10.1093/mnras/sty505}, 476, 4362

\bibitem[\protect\citeauthoryear{{Sorce}, {Courtois}, {Gottl{\"o}ber},
  {Hoffman}  \& {Tully}}{{Sorce} et~al.}{2014}]{2014MNRAS.437.3586S}
{Sorce} J.~G.,  {Courtois} H.~M.,  {Gottl{\"o}ber} S.,  {Hoffman} Y.,   {Tully}
  R.~B.,  2014, \mn@doi [\mnras] {10.1093/mnras/stt2153}, \href
  {https://ui.adsabs.harvard.edu/abs/2014MNRAS.437.3586S} {437, 3586}

\bibitem[\protect\citeauthoryear{Sorce et~al.,}{Sorce et~al.}{2016}]{Sorce2016}
Sorce J.~G.,  et~al., 2016, \mn@doi [\mnras] {10.1093/mnras/stv2407}, 455, 2078

\bibitem[\protect\citeauthoryear{{Sorce}, {Gottl{\"o}ber}  \& {Yepes}}{{Sorce}
  et~al.}{2020}]{2020MNRAS.496.5139S}
{Sorce} J.~G.,  {Gottl{\"o}ber} S.,   {Yepes} G.,  2020, \mn@doi [\mnras]
  {10.1093/mnras/staa1831}, \href
  {https://ui.adsabs.harvard.edu/abs/2020MNRAS.496.5139S} {496, 5139}

\bibitem[\protect\citeauthoryear{{Springel}}{{Springel}}{2005}]{2005MNRAS.364.1105S}
{Springel} V.,  2005, \mn@doi [\mnras] {10.1111/j.1365-2966.2005.09655.x},
  \href {https://ui.adsabs.harvard.edu/abs/2005MNRAS.364.1105S} {364, 1105}

\bibitem[\protect\citeauthoryear{{Stopyra}, {Peiris}, {Pontzen}, {Jasche}  \&
  {Natarajan}}{{Stopyra} et~al.}{2021}]{2021MNRAS.507.5425S}
{Stopyra} S.,  {Peiris} H.~V.,  {Pontzen} A.,  {Jasche} J.,   {Natarajan} P.,
  2021, \mn@doi [\mnras] {10.1093/mnras/stab2456}, \href
  {https://ui.adsabs.harvard.edu/abs/2021MNRAS.507.5425S} {507, 5425}

\bibitem[\protect\citeauthoryear{{Strauss} \& {Willick}}{{Strauss} \&
  {Willick}}{1995}]{1995PhR...261..271S}
{Strauss} M.~A.,  {Willick} J.~A.,  1995, \mn@doi [\physrep]
  {10.1016/0370-1573(95)00013-7}, \href
  {https://ui.adsabs.harvard.edu/abs/1995PhR...261..271S} {261, 271}

\bibitem[\protect\citeauthoryear{{Tsaprazi}, {Nguyen}, {Jasche}, {Schmidt}  \&
  {Lavaux}}{{Tsaprazi} et~al.}{2022}]{2022JCAP...08..003T}
{Tsaprazi} E.,  {Nguyen} N.-M.,  {Jasche} J.,  {Schmidt} F.,   {Lavaux} G.,
  2022, \mn@doi [\jcap] {10.1088/1475-7516/2022/08/003}, \href
  {https://ui.adsabs.harvard.edu/abs/2022JCAP...08..003T} {2022, 003}

\bibitem[\protect\citeauthoryear{{Tully}, {Shaya}, {Karachentsev}, {Courtois},
  {Kocevski}, {Rizzi}  \& {Peel}}{{Tully} et~al.}{2008}]{2008ApJ...676..184T}
{Tully} R.~B.,  {Shaya} E.~J.,  {Karachentsev} I.~D.,  {Courtois} H.~M.,
  {Kocevski} D.~D.,  {Rizzi} L.,   {Peel} A.,  2008, \mn@doi [\apj]
  {10.1086/527428}, \href
  {https://ui.adsabs.harvard.edu/abs/2008ApJ...676..184T} {676, 184}

\bibitem[\protect\citeauthoryear{{Tully} et~al.,}{{Tully}
  et~al.}{2013}]{2013AJ....146...86T}
{Tully} R.~B.,  et~al., 2013, \mn@doi [\aj] {10.1088/0004-6256/146/4/86}, \href
  {https://ui.adsabs.harvard.edu/abs/2013AJ....146...86T} {146, 86}

\bibitem[\protect\citeauthoryear{{Tully}, {Courtois}, {Hoffman}  \&
  {Pomar{\`e}de}}{{Tully} et~al.}{2014}]{2014Natur.513...71T}
{Tully} R.~B.,  {Courtois} H.,  {Hoffman} Y.,   {Pomar{\`e}de} D.,  2014,
  \mn@doi [\nat] {10.1038/nature13674}, \href
  {https://ui.adsabs.harvard.edu/abs/2014Natur.513...71T} {513, 71}

\bibitem[\protect\citeauthoryear{{Tully}, {Courtois}  \& {Sorce}}{{Tully}
  et~al.}{2016}]{2016AJ....152...50T}
{Tully} R.~B.,  {Courtois} H.~M.,   {Sorce} J.~G.,  2016, \mn@doi [\aj]
  {10.3847/0004-6256/152/2/50}, \href
  {https://ui.adsabs.harvard.edu/abs/2016AJ....152...50T} {152, 50}

\bibitem[\protect\citeauthoryear{{Tully} et~al.,}{{Tully}
  et~al.}{2022}]{2022arXiv220911238T}
{Tully} R.~B.,  et~al., 2022, arXiv e-prints, \href
  {https://ui.adsabs.harvard.edu/abs/2022arXiv220911238T} {p. arXiv:2209.11238}

\bibitem[\protect\citeauthoryear{{Tweed}, {Yang}, {Wang}, {Cui}, {Zhang}, {Li},
  {Jing}  \& {Mo}}{{Tweed} et~al.}{2017}]{2017ApJ...841...55T}
{Tweed} D.,  {Yang} X.,  {Wang} H.,  {Cui} W.,  {Zhang} Y.,  {Li} S.,  {Jing}
  Y.~P.,   {Mo} H.~J.,  2017, \mn@doi [\apj] {10.3847/1538-4357/aa6bf8}, \href
  {https://ui.adsabs.harvard.edu/abs/2017ApJ...841...55T} {841, 55}

\bibitem[\protect\citeauthoryear{{Valade}, {Hoffman}, {Libeskind}  \&
  {Graziani}}{{Valade} et~al.}{2022}]{2022MNRAS.513.5148V}
{Valade} A.,  {Hoffman} Y.,  {Libeskind} N.~I.,   {Graziani} R.,  2022, \mn@doi
  [\mnras] {10.1093/mnras/stac1244}, \href
  {https://ui.adsabs.harvard.edu/abs/2022MNRAS.513.5148V} {513, 5148}

\bibitem[\protect\citeauthoryear{{\lowercase{V}an de Weygaert} \&
  {Bertschinger}}{{\lowercase{V}an de Weygaert} \&
  {Bertschinger}}{1996}]{1996MNRAS.281...84V}
{\lowercase{V}an de Weygaert} R.,  {Bertschinger} E.,  1996, \mn@doi [\mnras]
  {10.1093/mnras/281.1.84}, \href
  {https://ui.adsabs.harvard.edu/abs/1996MNRAS.281...84V} {281, 84}

\bibitem[\protect\citeauthoryear{{Wang}, {Mo}, {Yang}, {Jing}  \& {Lin}}{{Wang}
  et~al.}{2014}]{2014ApJ...794...94W}
{Wang} H.,  {Mo} H.~J.,  {Yang} X.,  {Jing} Y.~P.,   {Lin} W.~P.,  2014,
  \mn@doi [\apj] {10.1088/0004-637X/794/1/94}, \href
  {https://ui.adsabs.harvard.edu/abs/2014ApJ...794...94W} {794, 94}

\bibitem[\protect\citeauthoryear{{Wang} et~al.,}{{Wang}
  et~al.}{2016}]{2016ApJ...831..164W}
{Wang} H.,  et~al., 2016, \mn@doi [\apj] {10.3847/0004-637X/831/2/164}, \href
  {https://ui.adsabs.harvard.edu/abs/2016ApJ...831..164W} {831, 164}

\bibitem[\protect\citeauthoryear{{Wang}, {Rooney}, {Feldman}  \&
  {Watkins}}{{Wang} et~al.}{2018}]{2018MNRAS.480.5332W}
{Wang} Y.,  {Rooney} C.,  {Feldman} H.~A.,   {Watkins} R.,  2018, \mn@doi
  [\mnras] {10.1093/mnras/sty2224}, \href
  {https://ui.adsabs.harvard.edu/abs/2018MNRAS.480.5332W} {480, 5332}

\bibitem[\protect\citeauthoryear{{Willick} \& {Strauss}}{{Willick} \&
  {Strauss}}{1998}]{1998ApJ...507...64W}
{Willick} J.~A.,  {Strauss} M.~A.,  1998, \mn@doi [\apj] {10.1086/306314},
  \href {https://ui.adsabs.harvard.edu/abs/1998ApJ...507...64W} {507, 64}

\bibitem[\protect\citeauthoryear{{Willick}, {Strauss}, {Dekel}  \&
  {Kolatt}}{{Willick} et~al.}{1997}]{1997ApJ...486..629W}
{Willick} J.~A.,  {Strauss} M.~A.,  {Dekel} A.,   {Kolatt} T.,  1997, \mn@doi
  [\apj] {10.1086/304551}, \href
  {https://ui.adsabs.harvard.edu/abs/1997ApJ...486..629W} {486, 629}

\bibitem[\protect\citeauthoryear{Yepes, Gottl{\"o}ber  \& Hoffman}{Yepes
  et~al.}{2014}]{YEPES20141}
Yepes G.,  Gottl{\"o}ber S.,   Hoffman Y.,  2014, \mn@doi [New Astronomy
  Reviews] {https://doi.org/10.1016/j.newar.2013.11.001}, 58, 1

\bibitem[\protect\citeauthoryear{{Zaroubi}, {Hoffman}, {Fisher}  \&
  {Lahav}}{{Zaroubi} et~al.}{1995}]{1995ApJ...449..446Z}
{Zaroubi} S.,  {Hoffman} Y.,  {Fisher} K.~B.,   {Lahav} O.,  1995, \mn@doi
  [\apj] {10.1086/176070}, \href
  {https://ui.adsabs.harvard.edu/abs/1995ApJ...449..446Z} {449, 446}

\bibitem[\protect\citeauthoryear{{Zaroubi}, {Zehavi}, {Dekel}, {Hoffman}  \&
  {Kolatt}}{{Zaroubi} et~al.}{1997}]{1997ApJ...486...21Z}
{Zaroubi} S.,  {Zehavi} I.,  {Dekel} A.,  {Hoffman} Y.,   {Kolatt} T.,  1997,
  \mn@doi [\apj] {10.1086/304481}, \href
  {https://ui.adsabs.harvard.edu/abs/1997ApJ...486...21Z} {486, 21}

\bibitem[\protect\citeauthoryear{{Zaroubi}, {Hoffman}  \& {Dekel}}{{Zaroubi}
  et~al.}{1999}]{1999ApJ...520..413Z}
{Zaroubi} S.,  {Hoffman} Y.,   {Dekel} A.,  1999, \mn@doi [\apj]
  {10.1086/307473}, \href
  {https://ui.adsabs.harvard.edu/abs/1999ApJ...520..413Z} {520, 413}

\bibitem[\protect\citeauthoryear{{Zel'dovich}}{{Zel'dovich}}{1970}]{1970A&A.....5...84Z}
{Zel'dovich} Y.~B.,  1970, \aap, \href
  {https://ui.adsabs.harvard.edu/abs/1970A&A.....5...84Z} {500, 13}

\makeatother
\end{thebibliography}



\appendix

\section{Posterior Gradient Derivation}
\label{AdjGrad}
Inference of the initial conditions performed using $\textsc{borg}$ usually consists of $\mathcal{O}(10^{6})$ or more free parameters. In order to sample efficiently very high-dimensional posterior spaces, \textsc{borg} makes use of an HMC sampling algorithm, which requires the gradient of the posterior distribution with respect to the parameters being inferred. The implementation of HMC in \textsc{borg} makes use of the tangent adjoint gradient of our forward data model, which is related to the gradient of the potential $\Psi \equiv - \ln \proba\left( \boldsymbol\delta^{\mathrm{I}} | \textbf{V}^{\mathrm{PV}} \right)$ with respect to the amplitude of the density contrast within a voxel, $\boldsymbol\delta^{\mathrm{I}}$. Indeed, our data model is composed of a sequence of elementary operators acting on any vector $\boldsymbol\delta^\mathrm{I}$ in parameter space, given as (see section \ref{sec::DataModel} and Figure \ref{fig:BHM})
\begin{equation}
\Psi(\boldsymbol\delta^\mathrm{I}) = (\Psi \circ V^\mathrm{r} \circ \textbf{V} \circ \textbf{v} \circ \boldsymbol\delta^\mathrm{F}) (\boldsymbol\delta^\mathrm{I}).
\label{modeloperators}
\end{equation}
The tangent adjoint gradient of the model is defined as the operator that acts upon any scalar $\tilde{\Psi}$ in `potential space' and returns the vector $\tilde{\boldsymbol\delta^\mathrm{I}}$ in parameter space, such that
\begin{equation}
\tilde{\boldsymbol\delta^\mathrm{I}} \equiv \left( \frac{\partial \boldsymbol\delta^{\mathrm{F}}}{\partial \boldsymbol\delta^{\mathrm{I}}} \right)^* \circ \left( \frac{\partial \textbf{v}}{\partial \boldsymbol\delta^{\mathrm{F}}} \right)^* \circ \left( \frac{\partial \textbf{V}}{\partial \textbf{v}} \right)^* \circ \left(\frac{\partial V^{\mathrm{r}}}{\partial \textbf{V}} \right)^* \circ \left( \frac{\partial \Psi}{\partial V^{\mathrm{r}}} \right)^* (\tilde{\Psi}),
\label{modeloperatorsderivatives}
\end{equation}
where $A^*$ denotes the Hermitian adjoint of operator $A$. This can be seen as the successive application, in reverse order, of the adjoint of the gradient of all the operators appearing in equation \eqref{modeloperators}, thus the name `tangent adjoint gradient'. The link between the tangent adjoint gradient operator and the gradient of the potential is obtained by the use of the chain rule of differentiation,
\begin{equation}
    \frac{\partial \Psi}{\partial \boldsymbol\delta^{\mathrm{I}}} = \frac{\partial \Psi}{\partial V^{\mathrm{r}}} \frac{\partial V^{\mathrm{r}}}{\partial \textbf{V}} \frac{\partial \textbf{V}}{\partial \textbf{v}} \frac{\partial \textbf{v}}{\partial \boldsymbol\delta^{\mathrm{F}}} \frac{\partial \boldsymbol\delta^{\mathrm{F}}}{\partial \boldsymbol\delta^{\mathrm{I}}} ,
\end{equation}
which involves all of the operators of which we need the adjoint in equation \eqref{modeloperatorsderivatives}. We derive explicit expressions for these operators and their adjoints in the following.

Using explicit indices, the gradient of the potential with respect to the value of the initial conditions in voxel $p$ is
\begin{equation}
    \frac{\partial \Psi}{\partial \boldsymbol\delta^{\mathrm{I}}_{p}} = \sum_{ n, \alpha, \beta, s, q} \frac{\partial \Psi}{\partial V^{\mathrm{r}}_{n}} \frac{\partial V^{\mathrm{r}}_{n}}{\partial \textbf{V}^{\alpha}_{n}} \frac{\partial \textbf{V}^{\alpha}_{n}}{\partial \textbf{v}^{\beta}_{s}} \frac{\partial \textbf{v}^{\beta}_{s}}{\partial \boldsymbol\delta^{\mathrm{F}}_{q}} \frac{\partial \boldsymbol\delta^{\mathrm{F}}_{q}}{\partial \boldsymbol\delta^{\mathrm{I}}_{p}} \,
\end{equation}
where $\mathbf{V}$ denotes the velocity field at tracer positions $\textbf{y}_n$, $\mathbf{v}$ denotes the velocity field on the grid at position $\textbf{x}_s$, superscript $\mathrm{r}$ denotes the line-of-sight component of the velocity field, the index $n$ runs over each peculiar velocity tracer, $\alpha$, $\beta$ denote Cartesian components of the velocity field, and the indices $s,q$ denote grid indices.

As we assume an independent Gaussian form for the likelihood of each tracer (equation \eqref{VIRBIUS_likelihood}), we have
\begin{align}
& \Psi(V^\mathrm{r}_n) = \frac{1}{2} \frac{\left[ V^\mathrm{PV}_n - V^{\mathrm{r}}_{n} \right]^{2}}{\sigma^{2}_{n}} + \mathrm{constant} , \nonumber\\
& \frac{\partial \Psi}{\partial V^{\mathrm{r}}_{n}} = - \frac{V^\mathrm{PV}_n - V^{\mathrm{r}}_n}{\sigma^{2}_{n}} ,
    \label{adjgrad1}
\end{align}
where $\sigma_{n}$ is the Gaussian error in the $n$th tracer line-of-sight peculiar velocity value.

From equation \eqref{LOS}, we see that
\begin{equation}
    \frac{\partial V^{\mathrm{r}}_{n}}{\partial \textbf{V}^{\alpha}_{n}} = \hat{y}^{\alpha}_{n} ,
    \label{adjgrad2}
\end{equation}
where $\hat{y}^{\alpha}_{n}$ is a Cartesian component of the unit direction vector at the $n$th tracer position.

From equation \eqref{CIC}, we see that
\begin{equation}
    \frac{\partial \textbf{V}^{\alpha}_{n}}{\partial \textbf{v}^{\beta}_{s}} = W\left(\textbf{x}_{s} - \textbf{y}_{n}\right) \delta_\mathrm{K}^{\alpha\beta},
    \label{adjgrad3}
\end{equation}
where $W$ denotes the CiC kernel function and $\delta_\mathrm{K}$ a Kronecker symbol. Equations (\ref{adjgrad1}), (\ref{adjgrad2}) and (\ref{adjgrad3}) involve real scalars, so the corresponding operators $\dfrac{\partial \Psi}{\partial V^{\mathrm{r}}}$, $\dfrac{\partial V^{\mathrm{r}}}{\partial \textbf{V}}$, and $\dfrac{\partial \textbf{V}}{\partial \textbf{v}}$, corresponding to real multiplications and summing over $n$ and $\alpha$, are self-adjoint. 

The velocity field on the grid is related to the $z=0$ density contrast by (see equation \eqref{discvfield})
\begin{equation}
     \textbf{v}^{\beta}_s = H f \sum_{tq} (\mathbfss{F}^{-1})_{st} \, \textbf{G}^{\beta}_{t} \,  \mathbfss{F}_{tq} \boldsymbol\delta^{\mathrm{F}}_{q},
     \label{adjgrad4a}
\end{equation}
where $\mathbfss{F}$ is a matrix denoting the three-dimensional DFT, given by its elements
\begin{equation}
    \mathbfss{F}_{ts} = \mathrm{e}^{-2 \mathrm{i}\pi \, t \cdot \left( \frac{s}{N} \right) },
    \label{DFT}
\end{equation}
where $t = (t^\lambda, t^\mu, t^\nu)$ is a three-dimensional index running on the Fourier space grid, $s = (s^\alpha, s^\beta, s^\gamma)$ is a three-dimensional index running on the configuration space grid, $t \cdot (s/N)$ denotes the scalar product, and the division $s/N = (s^\alpha/N, s^\beta/N, s^\gamma/N)$ is to be performed element-wise. Taking into account the normalisation factor for each dimension, the matrix corresponding to the inverse DFT has generic element
\begin{equation}
    (\mathbfss{F}^{-1})_{st} = \frac{1}{N^3} \mathrm{e}^{2 \mathrm{i}\pi \, t \cdot \left( \frac{s}{N} \right) } .
    \label{invDFT}
\end{equation}
$\boldsymbol\delta^{\mathrm{F}}_q$ is the final density contrast in voxel $q$, and the elements $\textbf{G}^{\beta}_{t}$ are defined by
\begin{equation}
    \textbf{G}^{\beta}_{t} \equiv -\mathrm{i} \frac{k^{\beta}_{t}}{|\textbf{k}_{t}|^{2}} ,
\end{equation}
where $\textbf{k}_t \equiv (2\pi t^\lambda/L, 2\pi t^\mu/L, 2\pi t^\nu/L)$ is the wavevector at the $t$th mode of the Fourier grid.

Taking the derivative in equation \eqref{adjgrad4a} gives
\begin{equation}
     \frac{\partial \textbf{v}^{\beta}_s}{\partial \boldsymbol\delta^\mathrm{F}_q} = Hf  \sum_{t} (\mathbfss{F}^{-1})_{st} \, \textbf{G}^{\beta}_{t} \, \mathbfss{F}_{tq} .
     \label{adjgrad4}
\end{equation}
Equations (\ref{DFT}) and (\ref{invDFT}) show that $(\mathbfss{F}^{*})_{ts} = N^3 (\mathbfss{F}^{-1})_{st}$ and $(\mathbfss{F}^{-1})^*_{st} = N^{-3} \mathbfss{F}_{ts}$. Therefore
\begin{equation}
    \left(\sum_{t} (\mathbfss{F}^{-1})_{st} \, \textbf{G}^{\beta}_{t} \, \mathbfss{F}_{tq} \right)^{*}  = - \sum_{t} \, \mathbfss{F}_{qt} \, \textbf{G}^{\beta}_{t} \, \mathbfss{F}_{ts} ,
\end{equation}
so that for any $\tilde{\textbf{v}}$,
\begin{equation}
\left[\left( \frac{\partial \textbf{v}}{\partial \boldsymbol\delta^\mathrm{F}} \right)^*(\tilde{\textbf{v}}) \right]^\beta_q  = - H f \sum_{ts} (\mathbfss{F}^{-1})_{qt} \, \textbf{G}^{\beta}_{t} \,  \mathbfss{F}_{ts} \tilde{\textbf{v}}_{s},
    \label{adjgrad4b}
\end{equation}
From equations \eqref{adjgrad4a} and \eqref{adjgrad4b}, we see that, up to a minus sign, $\left( \dfrac{\partial \textbf{v}}{\partial \boldsymbol\delta^{\mathrm{F}}} \right)^*$ is the same operator as $\textbf{v}$, i.e. a convolution by the kernel $H f \textbf{G}^\beta$ for each component $\beta$.

The final factor is the gradient of the operation that evolves the initial conditions to the $z=0$ density contrast. In the linear model (equation \eqref{EPTden}), we have simply
\begin{equation}
    \frac{\partial \boldsymbol\delta^{\mathrm{F}}_{q}}{\partial \boldsymbol\delta^{\mathrm{I}}_{p}} = D_{1}(t_0) \, \delta_\mathrm{K}^{qp} ,
    \label{adjgrad5}
\end{equation}
so that the operator $\dfrac{\partial \boldsymbol\delta^{\mathrm{F}}}{\partial \boldsymbol\delta^{\mathrm{I}}}$ is a multiplication by $D_{1}(t_0)$ and is self-adjoint. In the structure formation model case, we require the tangent adjoint gradient of the LPT evolution. The calculation is given in Appendix~D of \citet{2013MNRAS.432..894J} \citep[see also Appendix C in][]{2019A&A...625A..64J}. In this case, the tangent adjoint gradient is not limited to a multiplication by a real scalar.

\section{Sampler Warm-Up for Self-Consistency Test}
\label{ap::scburnin}
\begin{figure}
\centering
\includegraphics[width=\columnwidth]{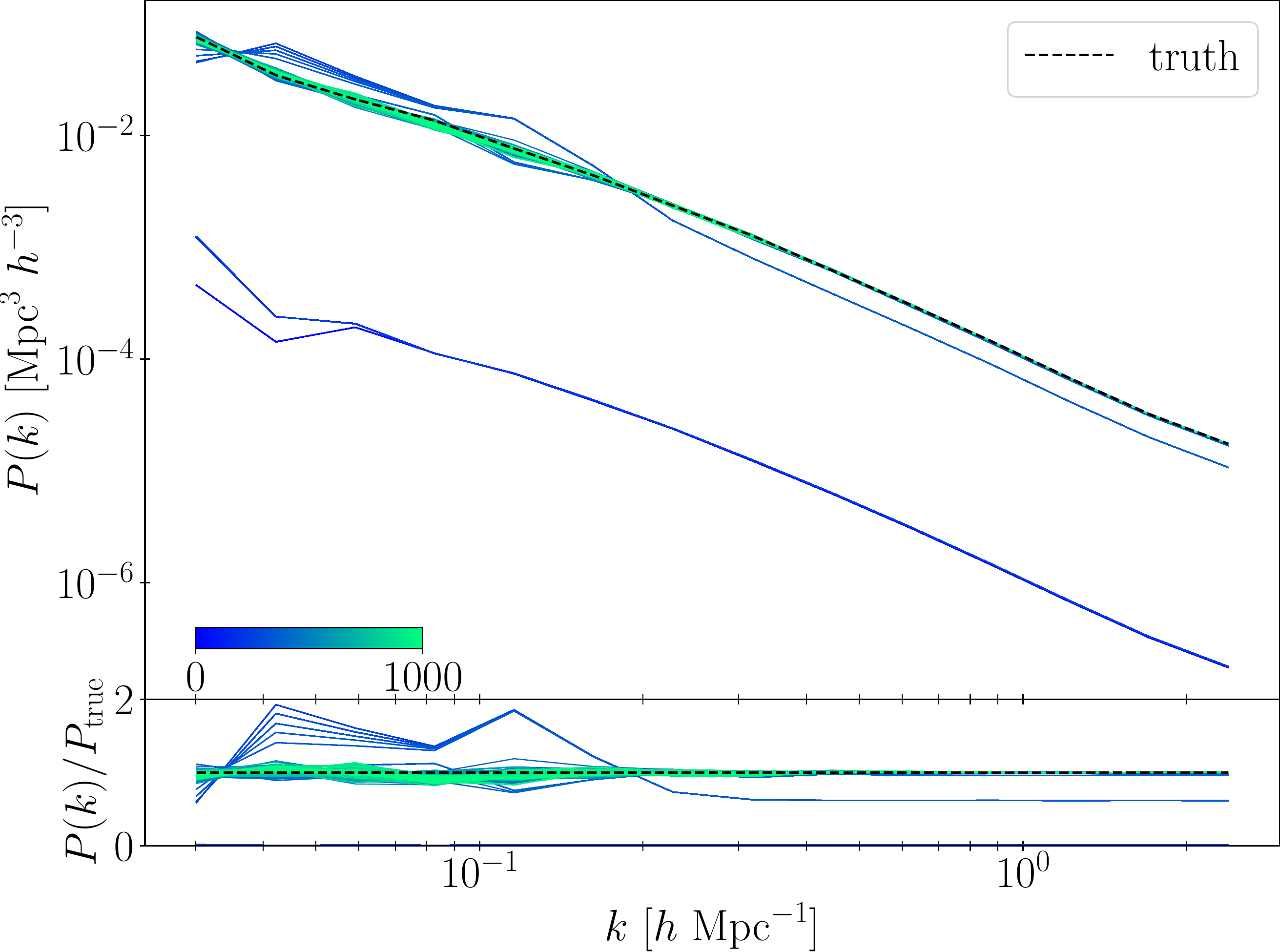}
\caption{\label{SCburn-in} Top: The posterior power spectrum of the initial conditions from the first 1000 samples of the chain, using the structure formation model and 10000 tracers. The samples are coloured according to their position in the chain (with the colouring given by the inset colour bar), and the spectrum of the true field is in black. Bottom: the ratio of the spectrum of each sample and the spectrum of the true field. The spectra of samples drift towards the spectrum of the true field, and begin to stably oscillate around it after around 500 samples. The sampler is then sampling the desired region of the posterior.}
\end{figure}

In order to investigate the warm-up behaviour of our Markov chain, we follow the approach used previously with $\textsc{borg}$ \citep[see e.g.][]{2013MNRAS.432..894J,2015JCAP...01..036J,2016MNRAS.455.3169L,2019A&A...625A..64J,2019arXiv190906396L}. We initialised the chain with an over-dispersed state such that the amplitude of density fluctuations was one-tenth of the predicted value from the $\Lambda$CDM case. We adjusted the step-size used by the HMC sampler, until samples were produced with an acceptance fraction of roughly 0.8. We then expect the Markov chain to drift towards the more likely regions of the posterior. This implies that the power spectra of samples of the posterior drifts towards the spectrum of the true field. In Figure \ref{SCburn-in}, we plot the spectra of the initial conditions from the first 1500 samples of the posterior, run using the structure formation model on the catalogue of 10000 tracers. As can be observed, the sequence of spectra drifts towards the spectrum of the true field and oscillates around it stably after around 1000 samples. This behaviour shows that the sampler has reached the desired region of the posterior. We performed the same test on the catalogue of 500 tracers, and then on both the catalogues using chains run with the linear model. All showed the same warm-up behaviour. 

\begin{figure}
\centering
\includegraphics[width=\columnwidth]{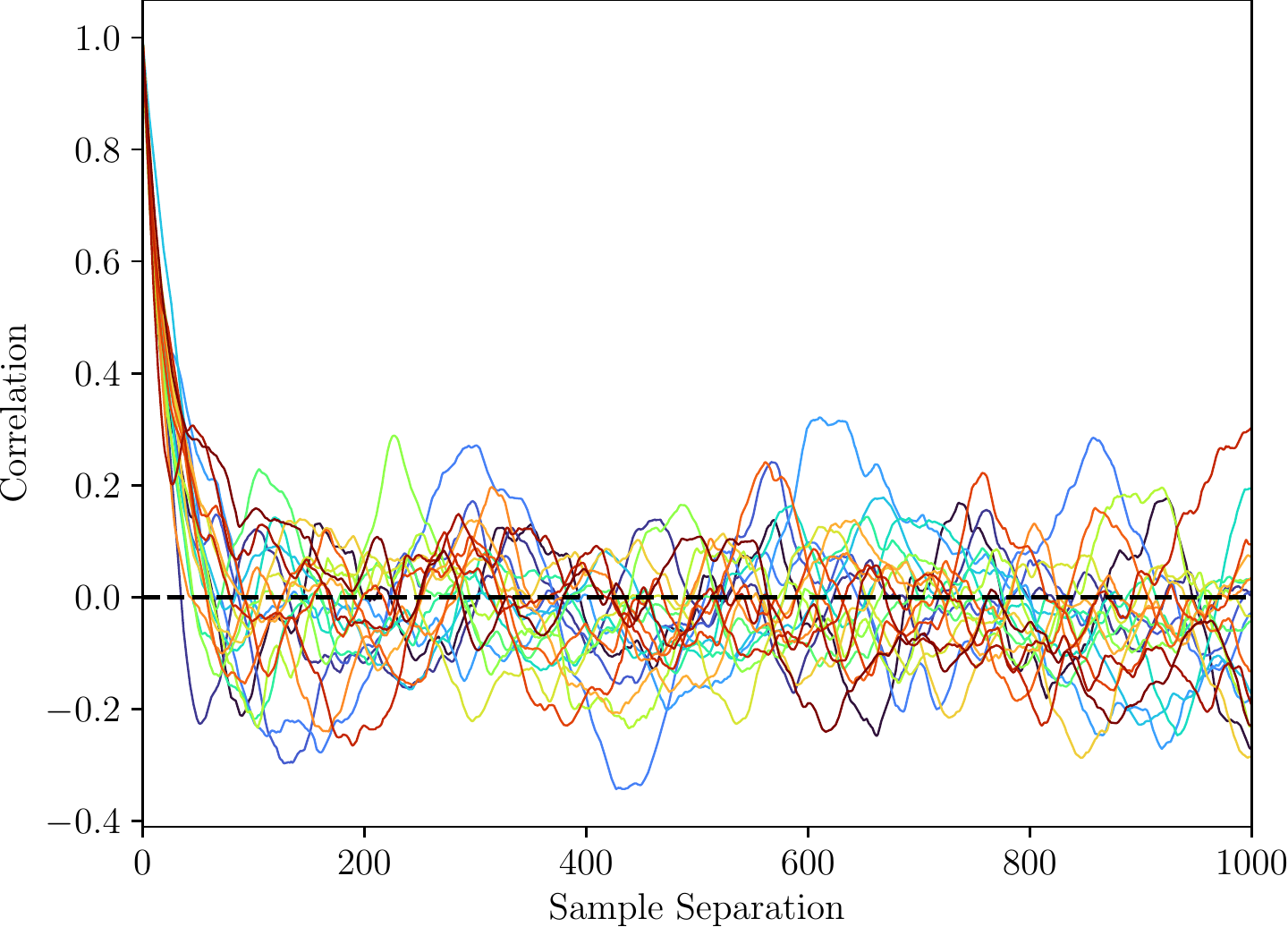}
\caption{\label{SCcorr} The autocorrelation of the inital density contrast within 20 voxels as a function of the separation between samples, using the structure formation model with 10000 tracers. The chain becomes decorrelated from the value in each voxel at the start of the chain after roughly 200 samples. This behaviour is typical of all voxels in the box.}
\end{figure}

Furthermore, in order to investigate the efficiency of our sampler, we considered the correlation length of our inference parameters, i.e. the values of the initial conditions in each voxel. In Figure \ref{SCcorr}, we plot the autocorrelation of the initial conditions within 20 voxels in the box, for 2000 samples of the posterior after warm-up, in the case of the chain run using the structure formation model with 10000 tracers. In each case, the value of the autocorrelation falls to zero after roughly 200 samples, implying that the chain becomes decorrelated from  its starting position after this many samples. This behaviour is typical for all voxels in the box, for both the chains with 500 and 10000 tracers.

\section{Quantitative Results of the Self-Consistency Tests with Sparse Data} \label{ap::scresults}
\begin{figure*}
\centering
\includegraphics[width=\textwidth]{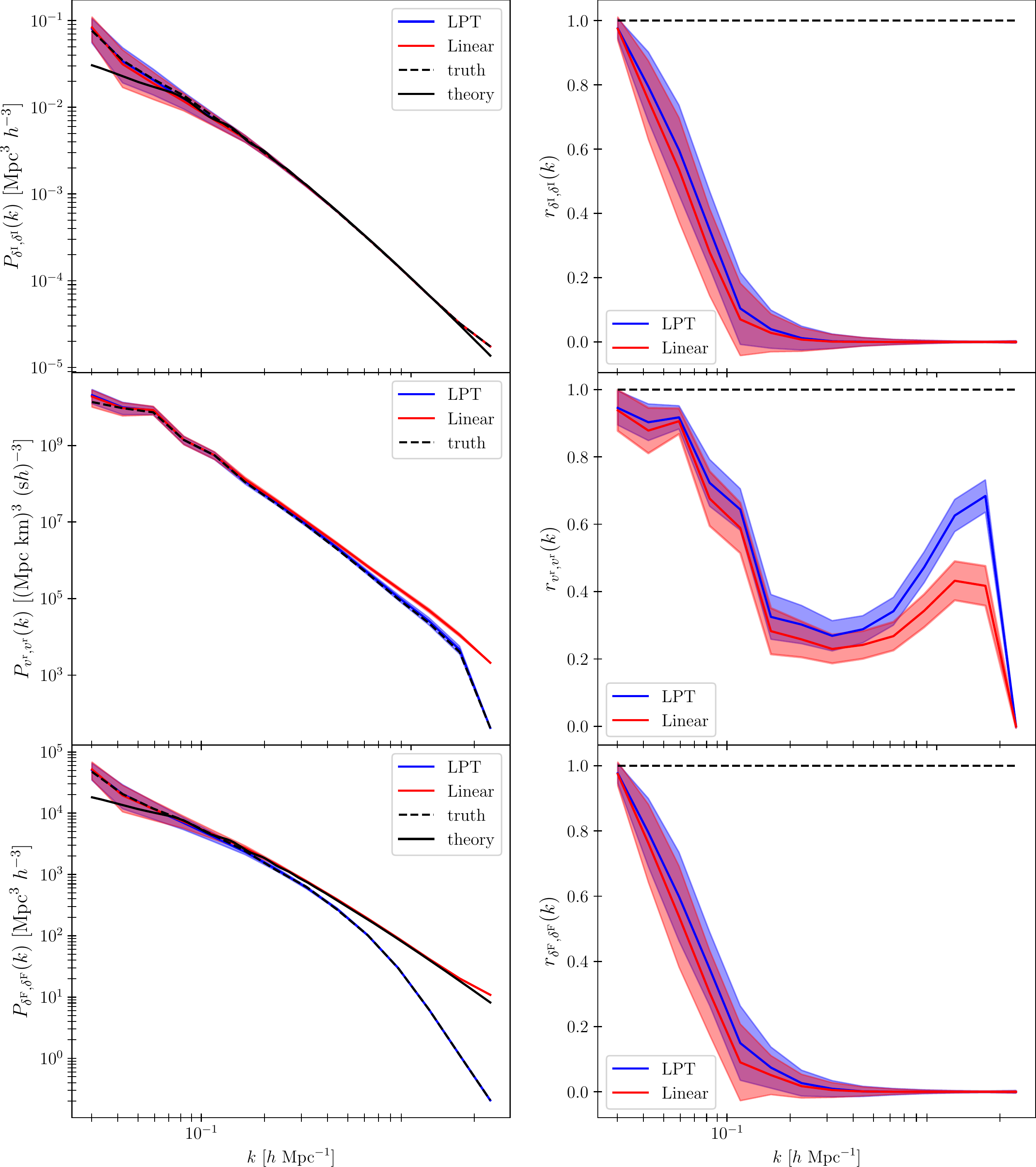} 
\caption{\label{SCspec500} Two-point statistics of field samples from chains run on the catalogue of 500 tracers. The left column shows power spectra and the right column shows the cross-correlation coefficient between samples and the corresponding true field. The first, second and third rows correspond to the initial density contrast $\delta^\mathrm{I}$, the $z=0$ line-of-sight velocity $v^\mathrm{r}$ and the $z=0$ density contrast $\delta^\mathrm{F}$, respectively. Two different data models are used in the inference: a model accounting for structure formation (`LPT', in blue), and a linear model (`Linear', in red). In each case, the chain mean and the $2\sigma$ credible region are shown, estimated from samples after warm-up. The dashed black line represents the true fields. For the initial conditions, the solid black line represents the theory prediction, and for the density contrast, the solid black line represents the initial conditions theory line linearly rescaled to $z=0$. The structure formation model better reproduces the two-point statistics of the true fields than the linear model, albeit with a greater variance than in the case of many tracers (Figure \ref{SCspec}). Therefore, our physical inference approach is robust in the limit of sparse data.}
\end{figure*}

We tested our structure formation model on a mock catalogue of 500 tracers made using the structure formation data model, in order to test whether the method is robust in the case of sparse data. In Figure \ref{SCspec500}, we show the power spectra and cross-correlation to the truth of samples produced using the structure formation and linear models. As can be seen in the top-left panel, both models produce samples of the initial conditions with power spectra consistent with the power spectrum of the true initial conditions. Compared to Figure \ref{SCspec}, the power spectra have a greater variance with fewer tracers. The middle-left and bottom-left panels, we plot the power spectrum of the line-of-sight velocity and density contrast respectively. We see that the structure formation fields reproduce the statistics of the true fields, while the linear model fields are biased high for $k \gtrsim 0.1$ $h$ Mpc$^{-1}$. Again, the variance of the power spectra is greater in the case of fewer tracers. In the right-hand column, we plot the cross-correlations between the chain samples and the true fields, as defined by equation  \eqref{cross-corr}. The structure formation model cross-correlations are greater than or equal to the linear model cross-correlations for all fields at all scales, showing that the structure formation model reconstructions have at least as good phase accuracy as the linear model reconstructions. We conclude that the structure formation model is robust in the limit of sparse data, as it is able to reconstruct the two-point statistics of the three fields.

\section{Sampler Warm-Up for Robustness to Model Mis-specification Tests}
\label{ap::hburnin}
\begin{figure}
\centering
\includegraphics[width=\columnwidth]{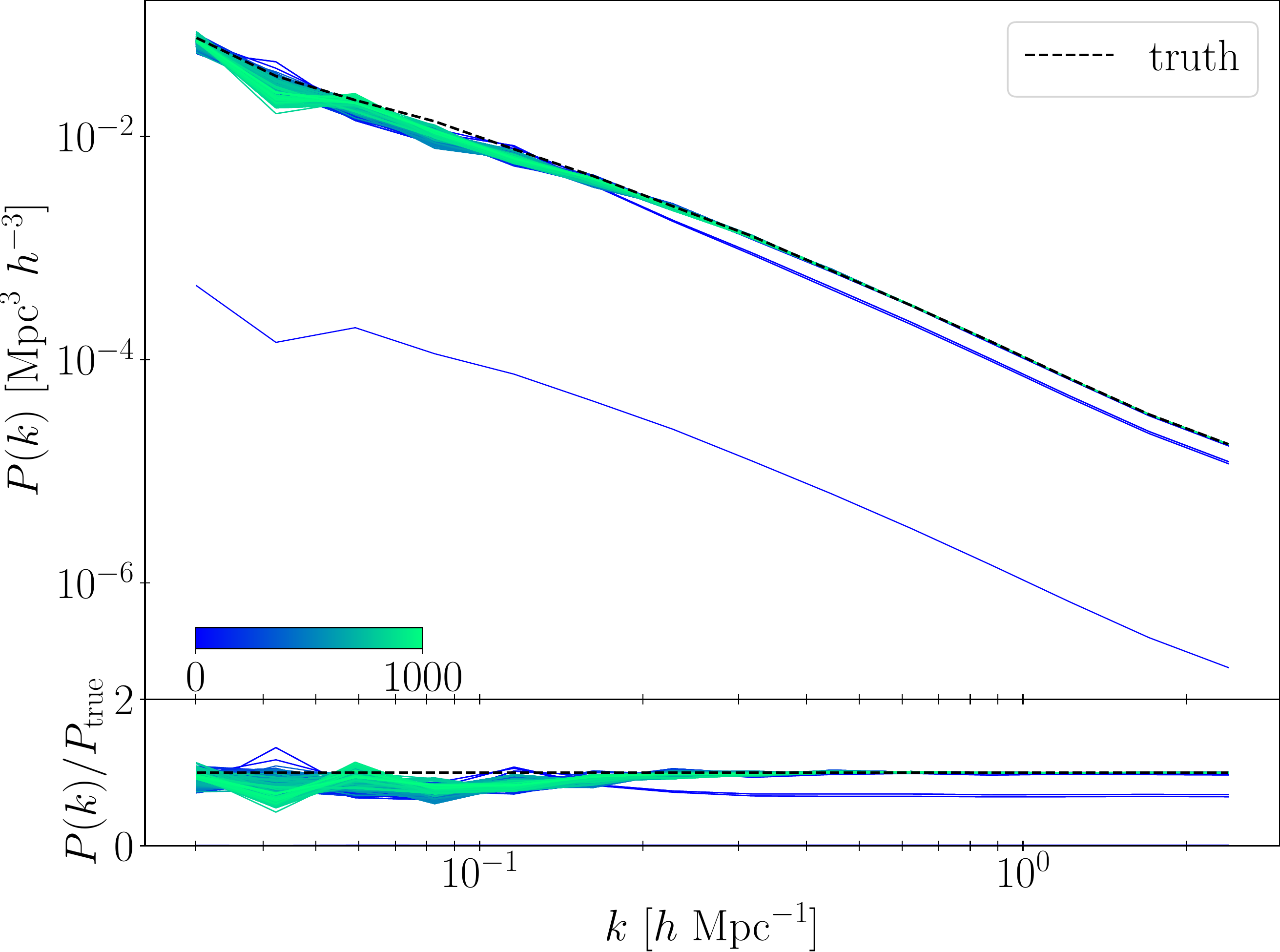}
\caption{\label{Hburn-in} Top: The posterior power spectrum of the initial conditions from the first 1000 samples of the chain, using the structure formation model and 10000 haloes as peculiar velocity tracers. The samples are coloured according to their position in the chain (with the colouring given by the inset colour bar), and the spectrum of the true field is in black. Bottom: the ratio of the spectrum of each sample and the spectrum of the true field. The spectra of samples drift towards the spectrum of the true field, and begin to stably oscillate around the small scale modes after around 500 samples.
}
\end{figure}

To test the warm-up behaviour of the sampler, we performed the same procedure as described in section \ref{ap::scburnin} for the self-consistency test. Figure \ref{Hburn-in} shows the power spectra of the initial conditions from the first 1000 samples of the chain, with the power spectrum of the initial conditions from which we generate the halo catalogue in black. Just as in the self-consistency case, the spectra of samples drift from the over-disperse starting point to oscillate around the true spectrum after a handful of samples, with the warm-up phase ending after roughly 500 samples.

We note that the large scale modes of the true field do not lie near the centre of the distribution of the corresponding samples; however, we consider the inference unbiased if the sample power spectra are consistent with the truth at the $2\sigma$ level, which is the case here.

\begin{figure}
\centering
\includegraphics[width=\columnwidth]{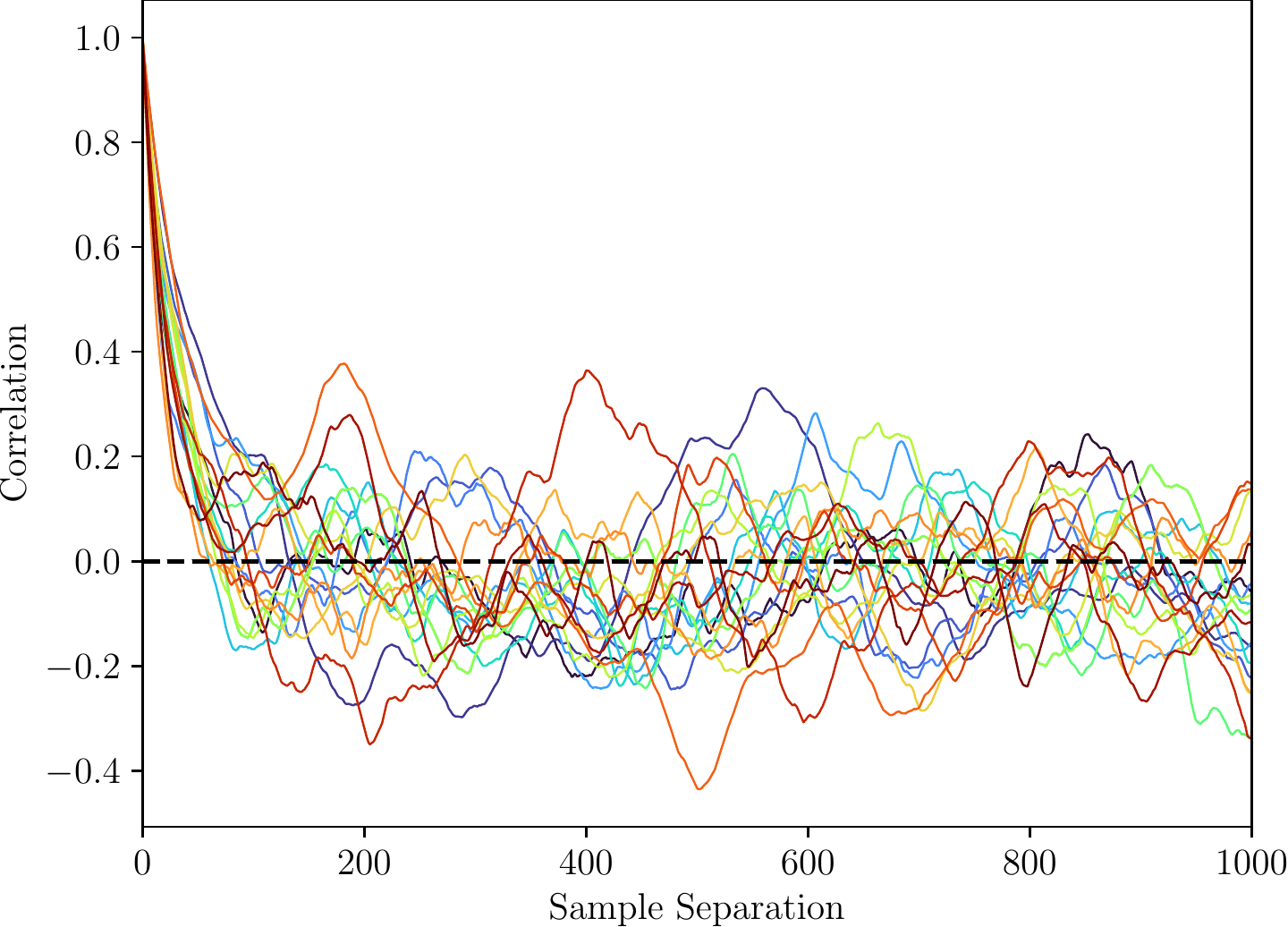}
\caption{\label{Hcorr} The autocorrelation of the inital density contrast within 20 voxels as a function of the sample number in the MCMC chain run using the structure formation model with 10000 mock haloes. The chain becomes decorrelated from the value in each voxel at the start of the chain after roughly 200 samples. This behaviour is typical of all voxels in the box.}
\end{figure}

Figure \ref{Hcorr} shows the autocorrelation of the initial conditions within 20 voxels in the box, for 1500 samples of the posterior after warm-up, in the case of the chain run using the structure formation model with 10000 mock haloes. Just like the self-consistency case, the autocorrelation falls to zero after roughly 200 samples for each voxel, implying that the chain becomes decorrelated from its starting position after this many samples. This behaviour is typical for all voxels in the box. 

\bsp	
\label{lastpage}
\end{document}